\documentclass{emulateapj}
\usepackage{mathptmx}

\def\vel{\upsilon}

\newcommand{\nv}{N{\,\sc v}}
\newcommand{\nvfull}{N{\,\sc v}~$\lambda1239$}
\newcommand{\nvdblt}{N{\,\sc v}~$\lambda\lambda1239,1242$}
\newcommand{\lya}{Ly$\alpha$}

\newcommand{\civ}{C{\,\sc iv}}
\newcommand{\civfull}{C{\,\sc iv}~$\lambda1548$}
\newcommand{\civdblt}{C{\,\sc iv}~$\lambda\lambda1548,1550$}
\newcommand{\ciii}{C{\,\sc iii}}
\newcommand{\ciiifull}{C{\,\sc iii}~$\lambda977$}

\newcommand{\siivfull}{Si{\,\sc iv}~$\lambda1394$}
\newcommand{\siiiifull}{Si{\,\sc iii}~$\lambda1207$}
\newcommand{\svi}{S{\,\sc vi}}
\newcommand{\svifull}{S{\,\sc vi}~$\lambda933$}
\newcommand{\kms}{km~s$^{-1}$}
\newcommand{\aox}{$\alpha_{\rm ox}$}
\newcommand{\ovi}{O{\,\sc vi}}
\newcommand{\ovifull}{O{\,\sc vi}~$\lambda1032$}
\newcommand{\ovidblt}{O{\,\sc vi}~$\lambda\lambda1032,1038$}

\newcommand{\noprint}[1]{}

\shorttitle{Intrinsic NAL Models for Three Quasars} 
\shortauthors{Wu et al.}
\slugcomment{To appear in the {\it Astrophysical Journal}}

\begin{document}
\title{The Physical Conditions of the Intrinsic N{\sc v} Narrow Absorption Line Systems of Three Quasars}

\author{Jian Wu\altaffilmark{1}, Jane C. Charlton\altaffilmark{1},
Toru Misawa\altaffilmark{1,2}, Michael Eracleous\altaffilmark{1,3}, \& 
Rajib Ganguly\altaffilmark{4}}
\email{jwu@astro.psu.edu}

\altaffiltext{1}{Department of Astronomy\&Astrophysics, the Pennsylvania
State University, 525 Davey Lab, University Park, PA, 16802}
\altaffiltext{2}{Current Address: Cosmic Radiation Laboratory RIKEN,
  2-1 Hirosawa, Wako City, Saitama, 351-0198, Japan}
\altaffiltext{3}{Center for Gravitational Wave Physics, The
  Pennsylvania State University, University Park, PA 16802}
\altaffiltext{4}{Current Address: Computer Science, Engineering, and
Physics Department, 213 Murchie Science Bld, University of Michigan,
Fling, MI 48502}

\begin{abstract}

  We employ detailed photoionization models to infer the physical
  conditions of intrinsic narrow absorption line systems found in high
  resolution spectra of three quasars at $z=2.6$--3.0. We focus on a
  family of intrinsic absorbers characterized by \nv\ lines that are
  strong relative to the \lya\ lines.  The inferred physical
  conditions are similar for the three intrinsic \nv\ absorbers, with
  metallicities greater than 10 times the solar value (assuming a
  solar abundance pattern), and with high ionization parameters
  ($\log{U}\sim0$).  Thus, we conclude that the unusual strength of
  the \nv\ lines results from a combination of partial coverage, a
  high ionization state, and high metallicity. We consider whether
  dilution of the absorption lines by flux from the broad-emission
  line region can lead us to overestimate the metallicities and we
  find that this is an unlikely possibility. The high abundances that
  we infer are not surprising in the context of scenarios in which
  metal enrichment takes place very early on in massive galaxies.  We
  estimate that the mass outflow rate in the absorbing gas (which is
  likely to have a filamentary structure) is less than a few ${\rm
    M_\odot~yr}^{-1}$ under the most optimistic assumptions, although
  it may be embedded in a much hotter, more massive outflow.

\end{abstract}


\section{Introduction}

{\it Intrinsic} absorption lines seen in the spectra of quasars are
thought to originate in ionized gas physically related with the quasar
``central engine'' \citep[e.g.,][]{ham04}.  They include resonance
lines that appear in the rest-frame UV band, such as N{\,\sc v},
C{\,\sc iv} and/or high ionization species that have absorption lines
and edges in the soft X-ray band, such as O{\,\sc vii} and O{\,\sc
viii} \citep[e.g.,][]{rey97,giustini10}. The intrinsic UV absorption
lines are usually blueshifted relative to the quasar emission lines
and they are believed to arise from gas that is outflowing from the
immediate vicinity of the central engine.  Lines that are blueshifted
by less than $5000\mbox{\ km\ s}^{-1}$ are conventionally regarded as
{\it associated} absorption lines (AALs). Many of these may indeed be
connected to broad emission-line region (BELR) \citep{gan01} and to
X-ray ``warm" absorbers \citep{mat95,bra00}. These absorption lines
thus trace the physical conditions and kinematics 
surrounding the central engine over a wide range of redshifts, 
and thus provide a diagnostic of quasar evolution.

Narrow intrinsic absorption lines (NALs) typically have widths
$\lesssim 500$~\kms\ so that their doublets can be separated at high
spectral resolution \citep[see][]{hamsab04}. This is in contrast to
broad absorption lines (BALs) whose width can reach 30,000~\kms\
\citep{tru06}.  Intrinsic NALs studies are complementary to 
studies of BALs, since they probe different regions in the quasar 
surroundings. Although the intrinsic NALs are more difficult to 
identify, several practical considerations make the derivation of
their physical conditions more straightforward than for BALs. 
\citep{ham99}. Because NALs are often unsaturated and
resolved, we can measure directly the NAL coverage fractions and
column densities of various ions.  In BAL systems, doublet transitions
are often self--blended, making such measurement more difficult.

Approximately 60\% of all quasars show evidence of outflows
\citep{mis07,gan08}. Although statistical methods can be used to
determine this, partial coverage of doublets and multiplets and time
variability analysis are the two most commonly used ways to
determine decisively if a particular NAL system is intrinsic.  These
signatures are only seen in low ionization transitions of intervening
absorbers, and only rarely in the case of small molecular clouds that 
could be smaller than the projected size of the background quasar
broad emission line region.  (e.g., Jones et al. 2010\nocite{jon10},
Ivanchik et al. 2010\nocite{iva10}). Partial coverage is described 
by the coverage fraction $C_{\rm f}$, which is
the fraction of photons from the background source that pass through
the absorber \citep{bhs97}.  It can be estimated using the residual
flux ratio of resonance doublets \citep[e.g.,][]{barsar97,gan99}.
Variability of the absorption lines could be caused by transverse
motion of the absorbing material or by changes in its ionization
state \citep{ham97,mis05}. Using UV spectra of $z\leqslant1.5$
quasars observed at two different epochs separated by 4--10 years with
the {\it Hubble Space Telescope}, \citet{wis04} concluded that a
minimum of $21\%$ of the AALs are variable. A similar conclusion was
reached for $z\sim 2$ quasars by \citet{nar04}.

\begin{deluxetable*}{lcccccccl}
\tablecolumns{6} 
\tabletypesize{\scriptsize} 
\tablecaption{Summary of Properties of Intrinsic NAL Systems \tablenotemark{a}\label{tab-sumsys}}
\tablewidth{0pc} 
\tablehead{
\colhead{}                                              &
\colhead{}                                              &
\colhead{}                                              &
\colhead{}                                              &
\colhead{}                                              &
\colhead{}                                              &
\colhead{}                                              &
\colhead{Spectral}                                      &
\colhead{}                                              \\
\colhead{}                                              &
\colhead{}                                              &
\colhead{$d_{\rm L}$\tablenotemark{b}}                  &
\colhead{}                                              &
\colhead{}                                              &
\colhead{}                                              &
\colhead{$\vel_{\rm ej}$\tablenotemark{f}}                 &
\colhead{Coverage\tablenotemark{g}}                     &
\colhead{}                                              \\
\colhead{Quasar}                                        &
\colhead{$z_{\rm em}$}                                  &
\colhead{(Gpc)}                                         &
\colhead{$\log L_{\rm bol}$\tablenotemark{c}}           &
\colhead{$\log Q_{\rm i}$\tablenotemark{d}}             &
\colhead{$z_{\rm abs}$\tablenotemark{e}}                &
\colhead{(km s$^{-1}$)}                                 &
\colhead{(\AA)}                                         &
\colhead{Transitions Detected}} 
\startdata
HE0130--4021 & 3.030 & 25.7 & 48.36 & 58.40 & 2.9739 &  4204 & 3612--6084 & Ly$\alpha$, Ly$\beta$, \ciii, \nv, \ovi\\
Q1009+2956   & 2.644 & 21.8 & 48.49 & 58.53 & 2.6495 & --453 & 3093--4623 & Ly$\alpha$, Ly$\beta$, \ciii, \svi, \nv, \ovi\\
HS1700+6416  & 2.722 & 22.6 & 48.98 & 58.97 & 2.7125 &   764 & 3726--6190 & Ly$\alpha$, Ly$\beta$, \civ, \nv, \ovi
\enddata
\tablenotetext{a}{Adopted from \citet{mis07}.}
\tablenotetext{b}{The luminosity distance, assuming $\Omega_\Lambda=0.7$, $\Omega_{\rm M}=0.3$, $h=0.7$.}
\tablenotetext{c}{The log of the bolometric luminosity (in erg~s$^{-1}$). See details in \S\ref{SED} of the text.}
\tablenotetext{d}{The log of the ionizing photon rate (in s$^{-1}$). See details in \S\ref{SED} of the text.}
\tablenotetext{e}{Absorption system redshift determined by finding the wavelength that divides the optical depth of the \nvfull\ line in half.}
\tablenotetext{f}{Ejection velocity relative to emission lines. A positive sign denotes a blueshift.}
\tablenotetext{g}{Wavelength range covered by the Keck spectrum.}
\end{deluxetable*}

Early work investigating metal abundances using narrow associated
absorption lines indicated supersolar metallicities, i.e., $Z\geqslant
Z_\odot$ \citep{pet94,tri96,ham97}. More recent work measures
metallicities consistent with the previously determined values, and in
some cases, it has provided a more stringent constraint. \citet{pap00}
analyzed a $z=1.207$ AAL system towards Q0122$+$0388 with the help of
photoionization models and found $Z\sim2Z_\odot$. \citet{fie05} found
that the metallicity must be at least $5Z_\odot$ in order to produce
the high-ionization absorption lines (O{\,\sc vi}, C{\,\sc iv}, and
N{\,\sc v}) of the narrow-line Seyfert galaxy Mrk 1044. \citet{gab06}
also found supersolar abundances by modelling the AALs of the quasar
J2233$-$606, with $[\mbox{C}/\mbox{H}]$,
$[\mbox{O}/\mbox{H}]\approx0.5$--0.9, and
$[\mbox{N}/\mbox{H}]\approx1.1-1.3$. High metallicity, up to
$3Z_\odot$, is suggested in the study of quasar PG 0935+417
\citep{ham04}. A study of the outflow from Mrk~279 by \citet{ara07}
has found an overabundance carbon, nitrogen, and oxygen relative to
the Sun by factors of $2.2\pm0.7$, $3.5\pm1.1$, and $1.6\pm0.8$,
respectively. Moreover, photoionization modeling of associated NALs
systems find metallicities ranging between a few and 20 times the
Solar value (e.g., Petitjean et al. 1994, Tripp, Lu \& Savage 1996,
D'Odorico et al. 2004) \nocite{pet94,tri96,dodorico04}.  Such high
metal abundances are also inferred from analysis of the broad emission
lines. For example, \citet{ham02} find that $\alpha$-elements, such as
C and O, are overabundant in the broad-emission line regions of
quasars by a factor of $\sim 3$ relative to the Sun.

Previous studies indicate that the AALs and BALs are located within
the scale of the host galaxy and suggest that the absorbers are
physically associated with and may originate from the quasar
outflow. For example, the strengths of the excited-state Si~{\sc
  ii}$^*$ AALs in the high resolution spectrum of \object{3C~191}
indicate a distance of $28$~kpc from the quasar assuming the gas is
photonionized \citep{ham01}. By studying the metastable excited levels
of Si~{\sc ii} and Si~{\sc ii}$^*$, \citet{dun10} determines the
distance to the outflow for attenuated and unattenuated SEDs of
$\sim6$~kpc or $\sim17$~kpc respectively. Variability analysis of a
\civ\ NAL complex towards quasar \object{HS~1603$+$3820} constrains an
upper limit on the distance of $r\leq6$~kpc from the continuum source
\citep{mis05}.

Although it is generally agreed upon that intrinsic absorption lines
are formed in an outflow from the central engine, there is no
consensus on the geometry or the driving force of such an outflow.  In
the model of \citet{mur95}, which was developed further by
\citet{pro00}, the wind is launched from the entire surface of the
accretion disk, with many denser, absorbing filaments distributed
throughout.  In this picture, the BALs appear at low latitudes (small
angles from the plane of the disk) when the line of sight passes
directly through the dense, high velocity wind. The NALs could form at
higher latitudes when the line of sight intercepts a dense filament in
the hot, low-density flow.  \citet{elv00}, instead, proposed a
funnel-shaped thin shell outflow model, in which the wind rises
vertically from a narrow range of radii on a disk in the BELR.  The
high-ionization UV NALs and the X-ray ``warm absorbers" are seen when
the disk is viewed at a low latitude, through the wind. When viewed
along the fast outflow, the full range of velocity is seen in
absorption with a large total column density, giving rise to a
BAL. Recent work indicates that the column densities measured in the
X-ray spectra of quasars hosting intrinsic NALs are considerably lower
than what the \citet{elv00} scenario posits \citep{mis08,chartas09},
which suggests a preference on the model of \citet{mur95}.  There is
also a suggestion that in some objects, the outflows are directed in
the polar direction, parallel to the axis of the accretion disk
\citep[see, for example,][]{ghosh07,zhou06,brother06}. In the context
of such a picture, and by analogy with equatorial wind models, the
absorbers giving rise to NALs could be detached from the dense part of
the outflow and become visible along different lines of sight.
Previous studies have shown that the absorbing clumps originate at
similar physical locations and are driven by radiative acceleration.
This is supported by the explanation of a double trough in the \civ\
BAL by the ``line-locking" effect (e.g., Arav et
al. 1995\nocite{ara}).  A striking case can be seen in \citet{sri02},
which reports an observation of a highly structured flow in which
distinct components have similar velocity separations.

\citet{mis07} identified two families of intrinsic absorption lines
among 39 intrinsic NAL systems. The ``strong N{\,\sc v}'' family is
characterized by strong, partially covered N{\,\sc v}, relatively weak
Ly$\alpha$ (less than twice the equivalent width of N{\,\sc v}),
occasionally detected C{\,\sc iv}, and O{\,\sc vi} lines (sometimes
even stronger than the N{\,\sc v} lines). The ``strong C{\,\sc iv}''
family, on the other hand, is characterized by strong, partially
covered C{\,\sc iv} doublets, strong, usually black Ly$\alpha$ lines,
and relatively weak or undetected N{\,\sc v}. A possible third family,
characterized by strong \ovi\ lines is discussed by \citet{ganpre}.
They primarily used partial covering to diagnose the intrinsic
nature of the systems, i.e., they found that the members
of the doublet could not be fit simultaneously
assuming full coverage. Depending on the geometries and locations
of the continuum source and broad emission line regions (and 
empirically on their positions relative to the emission lines)
the absorber can partially cover either or both \citep{gan99}.
There may be overlap between the \civ\ and \ovi\ categories as some
absorption systems can possess both strong \civ\ and
\ovi\ lines. Further observations covering both of these lines are
needed to investigate this issue.  These different types of NALs are
of interest because they may allow us to probe different regions of
the outflow. These three families of NALs may represent different
lines of sight through the outflow, which is also suggested by the
relations between the properties of UV NALs and the X-ray properties
of the quasars that display them \citep[e.g.,][]{chartas09}.

The question we address in this paper is the origin of the intrinsic
N{\,\sc v} NALs.  To this end we construct photoionization models for
the strong N{\,\sc v} absorption systems in the spectra of three
radio-quiet quasars from the HIRES/Keck sample studied by
\citet{mis07}. This sample contains $37$ optically bright quasars at
$z=2-4$. We choose these particular three quasars because their
spectra exhibit characteristics of the ``strong N{\,\sc v}" family. In
addition, these three systems offer many observational constraints
because many ions are covered in their spectra, and the \nv\
absorption lines have {\it relatively} simple profiles. Similar
systems, containing multiple absorption components spread over
thousands of \kms\ have been reported in the spectra of
RX~J1230.8$+$0115 \citep{gan03} and 3C~351 \citep{yua02}.

In Section~\ref{data}, we describe the absorption profiles of the
three intrinsic N{\,\sc v} NAL systems. In Section~\ref{method}, we
introduce our method for modelling the three systems using the Cloudy
photoionization code \citep{fer06}. The modelling results, in the form
of constraints on metallicity, ionization parameter, and volume number
density are presented in Section~\ref{results}. In
Section~\ref{discussion}, we discuss possible interpretations of our
results on the location of the absorbers in the quasar winds. We
present a summary and conclusion in Section~\ref{conclusion}. The
cosmology we use in this paper is $\Omega_\Lambda=0.7$, $\Omega_{\rm
  M}=0.3$, $h=0.7$, which leads to the luminosity distances listed in
Table~\ref{tab-sumsys}.

\section{Data} \label{data}

The Keck/HIRES spectra of our three intrinsic N{\,\sc v} systems are
described in \citet{mis07}. The spectral resolution is $R=37,500$, or
$\sim 7$~km s$^{-1}$.  Table~\ref{tab-sumsys} summarizes the basic
properties of the three absorption systems and of the quasars that
host them. The transitions listed are detected at a $5\sigma$
confidence level at the NAL redshift. The sign of the velocity of a
system is taken to be positive, if the line is blueshifted, i.e., if
the gas appears to be outflowing relative to the quasar; this is the
opposite convention from that adopted in \citet{mis07}. In the same
table, we also list the 4400~\AA\ flux densities \citep[taken
from][]{mis07}, as well as the bolometric luminosities and ionizing
photon rates (obtained as described in \S\ref{method}).

\begin{figure}
\centerline{\includegraphics[width=4in]{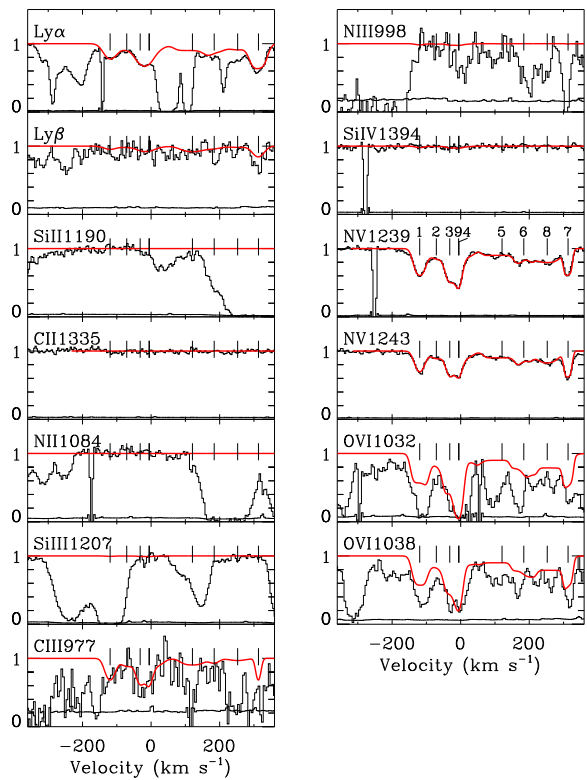}}
\vskip 0.5truein
\caption{Plots of absorption lines from all the transitions used as
  constraints for the $\vel_{\rm ej}=4204\;{\rm km\;s}^{-1}$ ($z_{\rm
    abs}=3.973915$) absorption system in the spectrum of
  HE0130$-$4021. Transitions are ordered by increasing ionization
  potential, except that Ly$\alpha$ and Ly$\beta$ (H{\sc i1026}) are
  always the first. The zero-point of the velocity scale is set by the
  bisector of the strongest blend of the N{\,\sc v} line (the bisector
  divides the optical depth of the blend in half).  Over-plotted (red)
  in each panel is the best theoretical profile we obtained by
  adjusting the photoionization model parameters (see
  Section~\ref{results} for details).  We mark each kinematic (Voigt)
  component using a tick mark; the component numbers are printed in
  the \nvfull\ panel. The corrsponding componets for different
  transitions have the same partial coverage factor (see
  Section~\ref{method}).  The corresponding components for different
  transitions have the same partial coverage factor (see
  Section~\ref{method}), which are listed in
  Table~\ref{tab-nvfit}.\label{fig-q0130-sys}\smallskip}
\end{figure}

\begin{figure}
\centerline{\includegraphics[width=4in]{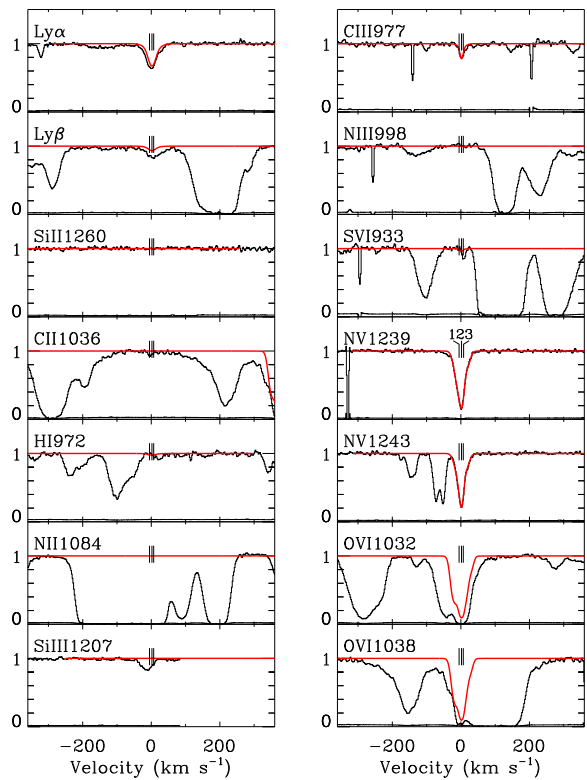}}
\vskip 0.5truein
\caption{Same as Fig.~\ref{fig-q0130-sys} but for the
$\vel_{\rm ej}=-453\;{\rm km\;s}^{-1}$ ($z_{\rm abs}=2.649510$)
absorption system in the spectrum of
Q1009$+$2956. \label{fig-q1009-sys}}
\end{figure}

\begin{figure}
\centerline{\includegraphics[width=4in]{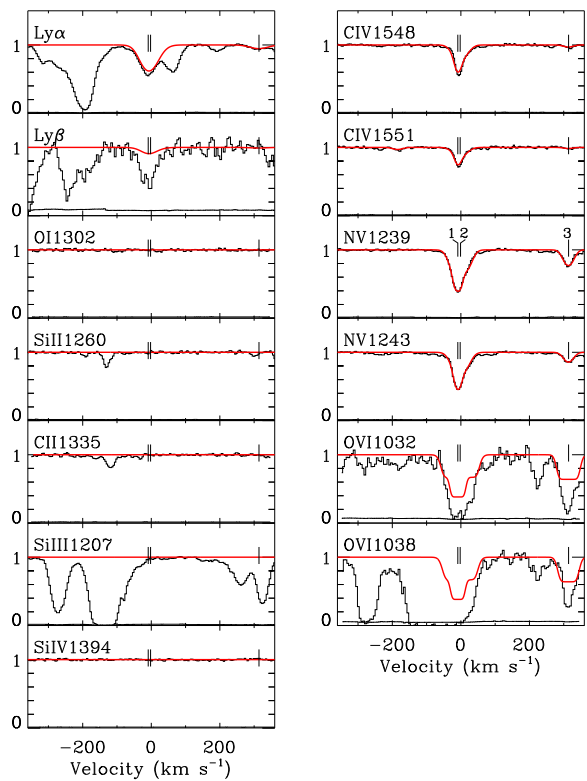}}
\vskip 0.5truein
\caption{Same as Fig.~\ref{fig-q0130-sys} but for the
$\vel_{\rm ej}=764\;{\rm km\;s}^{-1}$ ($z_{\rm abs}=2.7125$) absorption
system in the spectrum of HS1700$+$6416.\label{fig-q1700-sys}}
\end{figure}

Figures~\ref{fig-q0130-sys}, \ref{fig-q1009-sys}, and
\ref{fig-q1700-sys} present, for each system, absorption profiles for
transitions which are used as modelling constraints. An example of the
best models we have found (discussed in later sections) is also shown
in each figure. The velocity of an entire system is defined by the
optical depth-weighted center of the strongest member of the N{\,\sc v} 
doublet. We describe each system below.

\begin{deluxetable*}{lccrccc}
\tablecolumns{7} 
\tablewidth{0pc}
\tablecaption{Results of Decomposition of N{\,\sc v} Doublet Profiles \tablenotemark{a}\label{tab-nvfit}} 
\tablehead{
\colhead{}                                     &
\colhead{Kinematic}                            &
\colhead{$\vel_{\rm ej}\;$\tablenotemark{b}}  &
\colhead{$\vel_{\rm rel}\;$\tablenotemark{c}} &
\colhead{}                                     &
\colhead{$b\;$\tablenotemark{e}}           &
\colhead{}                                    \\
\colhead{Quasar}                               &
\colhead{Component}                            &
\colhead{(km s$^{-1}$)}                &
\colhead{(km s$^{-1}$)}                &
\colhead{$\log(N_{\rm N\,V}/{\rm cm}^{-2})\;$\tablenotemark{d}} &
\colhead{(km s$^{-1}$)}                &
\colhead{$C_{\rm f}\;$\tablenotemark{f}} \\
}
\startdata
HE0130$-$4021 & 1 &  4324  & $-120$ ~~ & $14.18\pm0.04$ & $16.6\pm0.9$  & $0.43\pm0.03$\\
              & 2 &  4275  & $-71$ ~~  & $14.2\pm0.1$   & $20\pm8$      & $0.16\pm0.04$\\
              & 3 &  4236  & $-32$ ~~  & $13.9\pm0.1$   & $13\pm2$      & $0.49\pm0.08$\\
              & 4 &  4209  & $-5$ ~~   & $14.8\pm0.3$   & $32\pm5$      & $0.13\pm0.03$\\
              & 5 &  4084  & $+121$ ~~ & $15.9\pm0.5$   & $28\pm5$      & $0.10\pm0.02$\\
              & 6 &  4020  & $+184$ ~~ & $15.0\pm0.3$   & $17\pm2$      & $0.13\pm0.03$\\
              & 7 &  3891  & $+313$ ~~ & $15.2\pm0.6$   & $6\pm1$       & $0.37\pm0.02$\\
              & 8 &  3952  & $+252$ ~~ & $14.62\pm0.06$ & $34\pm3$      & $0.21\pm0.02$\\
              & 9 &  4210  & $-6$ ~~   & $13.51\pm0.01$ & $12.6\pm0.3$  & $1.00$\tablenotemark{g}\\
\noalign{\vskip 10pt}
Q1009$+$2956  & 1 & $-453$ & $-5$ ~~   & $13.70\pm0.29$ & $14.3\pm1.7$  & $0.62\pm0.11$\\
              & 2 & $-455$ & $+2$ ~~   & $14.79\pm1.48$ & $3.40\pm1.90$ & $0.70\pm0.06$\\
              & 3 & $-461$ & $+7$ ~~   & $14.63\pm0.28$ & $13.0\pm2.1$  & $0.22\pm0.05$\\
\noalign{\vskip 10pt}
HS1700$+$6416 & 1 &  $771$ & $-9$ ~~   & $14.18\pm0.05$ & $11.0\pm0.5$  & $0.43\pm0.02$\\
              & 2 &  $766$ & $+2$ ~~   & $14.62\pm0.04$ & $25.6\pm0.5$  & $0.33\pm0.02$\\
              & 3 &  $450$ & $+314$ ~~ & $13.8\pm0.1$   & $18.4\pm0.8$  & $0.36^{+0.07}_{-0.06}$ 
\enddata
\tablenotetext{a}{The errors are a combination of the continuum
  fitting errors and Voigt profile fitting errors.}
\tablenotetext{b}{Ejection velocity relative to the emission lines. A positive sign denotes a blueshift.}
\tablenotetext{c}{Velocity offset relative to the redshift of the entire 
  absorption line complex. The sign of the velocity is set to be positive, if 
  the lines are redshifted from the system center.}
\tablenotetext{d}{The \nv\ column density inferred from the Voigt profile fit.}
\tablenotetext{e}{Doppler parameter.}
\tablenotetext{f}{Effective coverage fraction.}
\tablenotetext{g}{There is no error bar for this entry because our
  fitting yields $C_{\rm f}>1$, which is unphysical. Thus, we set
  $C_{\rm f} = 1.00$ when we fit this component.}
\end{deluxetable*}

\begin{description}

\item {\it HE0130$-$4021} ($z_{\rm abs}=2.973915$; 
  Fig.~\ref{fig-q0130-sys}): The system is found $\sim 4000$~\kms\
  blueward of the quasar emission redshift, just within the associated
  region. This system is kinematically more complex than the other two,
  with 9 components in each N{\,\sc v} line, spreading over $\sim
  500$~\kms. All but one of these components have small coverage
  fractions ($C_{\rm f} < 0.5$; see Table~\ref{tab-nvfit}). The strong
  features in the Ly$\alpha$ panel of Figure~\ref{fig-q0130-sys} at
  $\sim -280$, $50$, $110$, and $210$~\kms\ cannot be Ly$\alpha$
  components because the corresponding Ly$\beta$ components are not
  detected.  Thus only portions of the Ly$\alpha$ profile can be used
  as constraints, but it is clear that the absorption in Ly$\alpha$ is
  quite weak.  The O{\,\sc vi} doublet appears to be detected in many
  of the same components as N{\,\sc v}, though inconsistencies between
  O{\,\sc vi}~$\lambda1032$ and O{\sc vi}~$\lambda1038$ betray a
  number of blends.  For this system the alignment of kinematic
  components of different transitions is convincing and it seems at
  least some of the O{\sc vi} absorption is real.  We can use the
  observed absorption at the positions of \ovidblt\ as an upper limit.

\item {\it Q1009$+$2956} ($z_{\rm abs}=2.649510$; 
  Fig.~\ref{fig-q1009-sys}): This system is $\sim450$~\kms\ redward of
  the quasar emission redshift.  The \nvdblt\ doublet of this system
  exhibits partial coverage in each of the three blended kinematic
  (Voigt) components required to fit its asymmetric profile, with the
  strongest component having $C_{\rm f} = 0.70\pm0.06$
  (Fig.~\ref{fig-q1009-sys} and Table~\ref{tab-nvfit}).  Ly$\alpha$
  and Ly$\beta$ are detected, but are unusually weak compared to 
  N{\,\sc v}.  The ratio of Ly$\alpha$ to Ly$\beta$ is reasonable, but
  differences in the shapes of these lines, particularly the presence
  of a wing on the red side of the Ly$\beta$ profile, suggest a
  possible blend with another line. We will thus weight the Ly$\alpha$
  line more heavily as a constraint.  Narrow \ciiifull\ and \svifull\
  lines appear to be detected, albeit with some uncertainty because of
  their location in the Ly$\alpha$ forest. The red member of S{\,\sc
    vi~$\lambda\lambda 933,945$} doublet is heavily blended and thus
  cannot be used to judge whether this doublet is truly detected or
  not. The feature at the position of \siiiifull\ is not aligned with
  the \ciiifull\ and \nvfull\ profiles and is too broad, so we believe
  that we only have an upper limit on \siiiifull\ absorption from this
  system. There is absorption at the positions of both members of the
  \ovidblt\ doublet, however the minima are not aligned with each
  other or with the N{\,\sc v} profile. The \ovifull\ is too broad
  relative to O{\sc vi}~$\lambda$1038, and both transitions are
  clearly affected by blends.  The limits on O{\,\sc vi} for this
  system are not restrictive constraints for our models.

\item {\it HS1700$+$6416} ($z_{\rm abs}=2.7125$;
  Fig.~\ref{fig-q1700-sys}): This system is $\sim 750$~km~s$^{-1}$
  blueward of the quasar emission redshift. The \nvdblt\ and \civdblt\
  doublets are detected in this system, as well as Ly$\alpha$ and
  Ly$\beta$ (Fig.~\ref{fig-q1700-sys}).  The same system was presented
  in \citet{bhs97}, who found it to vary significantly on a time scale
  of $\sim6.5$ months in its rest-frame.  The central absorption is
  detected at $z\sim2.7125$ ($\vel_{\rm rel}\sim0$~km~s$^{-1}$) and a
  weaker component is found at $\sim325$~km s$^{-1}$.  The central
  component shows an asymmetry, indicating the need for two Voigt
  components for an adequate fit.  The feature at $\sim
  70$~km~s$^{-1}$ in the Ly$\alpha$ panel cannot be Ly$\alpha$
  belonging to this absorption system, since no counterpart is
  detected in the Ly$\beta$ panel.  All three components of the
  N{\,\sc v} profile show partial coverage, with coverage fractions
  between $0.33$ and $0.43$ (Table~\ref{tab-nvfit}). The Ly$\beta$
  component at $0$~km s$^{-1}$ is too strong relative to Ly$\alpha$,
  indicating a blend that affects the Ly$\beta$ so that it can only be
  used as an upper limit.  If saturated (and affected by partial
  coverage), Ly$\beta$ could have an equivalent width equal to that of
  Ly$\alpha$, but it can never have a larger equivalent width.  The
  O{\,\sc vi} doublet lies in the Ly$\alpha$ forest, such that it is
  hard to evaluate whether it is detected for this system or not.  If
  all of the detected absorption, either at $0\mbox{\ km\ s}^{-1}$ or
  at $325$~km~s$^{-1}$, is due to O{\,\sc vi}, the absorber would have
  a much larger coverage fraction than the N{\,\sc v}, C{\,\sc iv}, or
  Ly$\alpha$ absorbers at these same velocities. In our modelling we
  will take the absorption lines coincident with O{\,\sc vi} as an
  upper limit, but we note that the minima of the absorption at $\sim
  325$~km~s$^{-1}$ in both members of the O{\,\sc vi} doublet are at
  the same velocity as those in the N{\,\sc v}. Because of this, we
  will later consider whether viable single phase models can also
  explain this possible O{\,\sc vi} absorption.

\end{description}

 Transitions appearing in Figures~\ref{fig-q0130-sys},
\ref{fig-q1009-sys}, and \ref{fig-q1700-sys}, but not described above,
provide limits for model constraints, either because they are not
detected or because they thought to be affected by blends.

\section{Photoionization Modeling} \label{method}

Our calculations were performed with version 07.02.01 of Cloudy, last
described by \citet{fer98}.  Our modeling assumes a series of
plane-parallel slabs of gas (clouds) exposed to the ionizing
continuum from the central engine.  
The gas within each absorbing cloud is assumed to have a uniform density,
metallicity and abundance pattern. In most cases, we assume a solar
abundance pattern \citep{hol01}, so that for {\it every} element $i$,
$A_i/A_{i,\odot}=Z/Z_\odot$ where $A_i=n_i/n_{\rm H}$. When an
adequate fit cannot be achieved using the solar abundance pattern, we
consider deviations.  The intensity of the ionizing continuum 
is parameterized by the ionization parameter,
\begin{equation}
\label{eq3-u}
U=\frac{n_\gamma}{n_{\rm H}}=\frac{1}{4\pi r^2\, c \, n_{\rm H}}
\int^{\epsilon_2}_{\epsilon_1}\frac{L_\epsilon}{\epsilon}\,d\epsilon,
\end{equation}
where $r$ is the distance of the illuminated face of the cloud from
the continuum source, $\epsilon_1$ is the lowest energy
required to photoionize the gas, i.e., $\epsilon_1=1$~Ry, and
$\epsilon_2$ is the high energy cutoff of the SED. In our model, we
take $\epsilon_2=7.354\times10^6$~Ry. Each of our models are specified
by $U$ and $n_{\rm H}$, which (given the quasar luminosity and spectral
shape) corresponds to a given $r$. We also assume that all the
ions within a cloud have the same coverage fraction, and that the gas
cloud is in a state of thermal equilibrium, so it is parameterized by
a single electron temperature, $T_e$.  The assumption of the same
coverage fraction for different lines is likely to be valid for those
from ions with similar ionization states, particularly \civ\ and
\nv\ which are the most important constraints for out
models. \lya\ absorption may also arise from additional regions, but
in any case will provide a lower limit on metallicity. In our favored
models, the clouds are optically thin to the incident continuum so
that the incident SED does not change
after the ionizing photons pass through an absorbing cloud.

We have measured the Doppler $b$ parameters, column densities and
coverage fractions of the N{\,\sc v} doublets, as we describe below,
and optimized on these values (i.e., we required that Cloudy
models produce them). We choose N{\,\sc v} as the transition on which
we optimize because it is located in a relatively ``clean'' spectral
region where there are few blends. This allows us to determine its
column density, Doppler parameter, and coverage fraction via Voigt
profile fitting. The code AUTOVP \citep{dav97} is used to
derive an initial solution, and then MINFIT is used to determine the
minimum number of components that produce an adequate fit.
The goal of the modeling exercise is to reproduce the observed 
absorption profiles for all other ions by adjusting three physical 
parameters: metallicity, $Z/Z_\odot$, 
ionization parameter, $U$, and hydrogen number density, $n_{\rm H}$.

\subsection{Voigt Profile Fitting of the N{\,\sc v} Doublets}

\begin{figure}
\rightline{\includegraphics[width=4.0in]{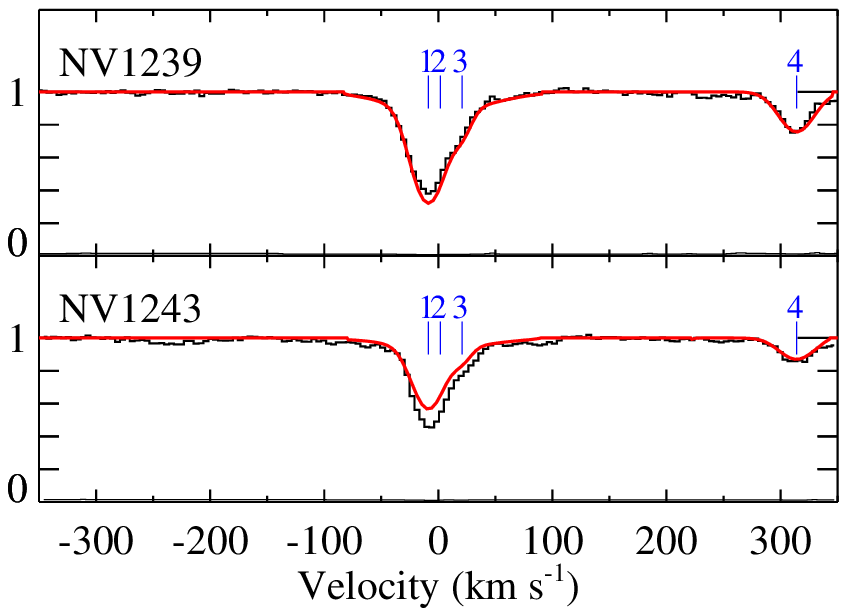}}
 \vspace{1.5cm}
\rightline{\includegraphics[width=4.0in]{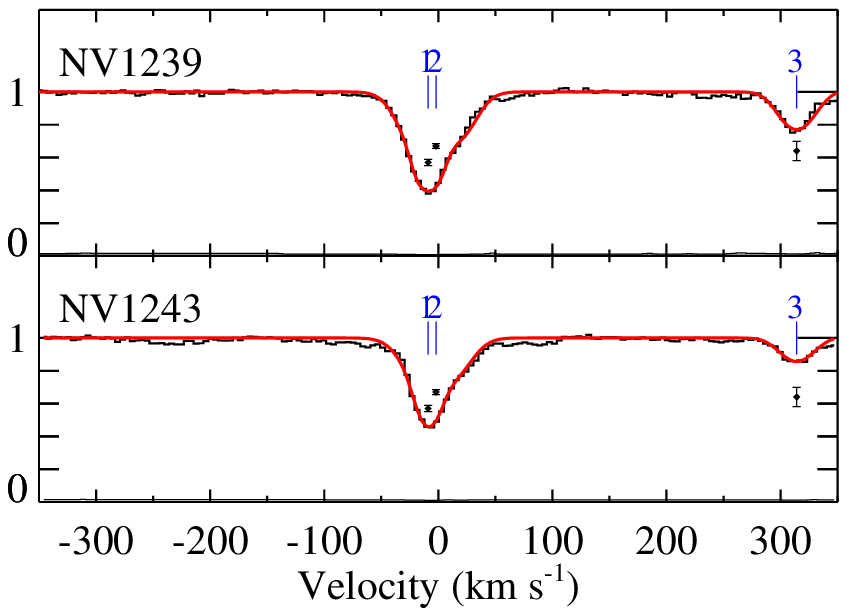}}
 \vspace{1cm}
\caption{Voigt profile fits to the \nvdblt\ doublet in the $\vel_{\rm
    ej}=764\;{\rm km\;s}^{-1}$ ($z_{\rm abs}=2.7125$) system in the
  spectrum of HS1700$+$6416. The velocity zero point is defined as in
  Fig.~\ref{fig-q0130-sys}. The fitting process minimizes the number
  of kinematic components needed for an adequate representation of the
  profile. Vertical ticks mark the component centers. In the upper
  pair of panels the coverage factors of all components are fixed to
  be unity; in the lower pair of panels, the coverage factors of all
  components are treated as free parameters. Black dots with error
  bars represent the value of ($1-C_{\rm f}$) for each
  component. Clearly, $C_{\rm f}=1$ does not provide an adequate fit,
  so that partial coverage is required. \label{fig-q1700-cf}}
\end{figure}

The approach of the MINFIT code is to first ``overfit'' the system
using many Voigt components and then to reject components that do not
improve the fits at a confidence level above $95\%$. This fitting
technique has been used extensively in studies of intervening Mg{\,\sc
  ii} systems (e.g., Ding et al. 2003; Ding, Charlton \& Churchill
2005; Lynch, Charlton \& Kim 2006\nocite{din03,din05,lyn06}). Table
\ref{tab-nvfit} lists the resulting fitting parameters.  It is
essential to include the coverage fractions as free parameters in the
fitting process for these systems since $C_{\rm f}=1$ does not provide
acceptable fits to these doublets within the observed uncertainties
(Fig.~\ref{fig-q1700-cf}; Misawa et al.  2007)\nocite{mis07}.

\subsection{Specification of the Ionizing Continuum} \label{SED}

Although multi-band
photometry and/or spectra of these three quasars are not available
over the entire spectral range, some observational data are available
to place constraints on the input SED.  We consider three possible
quasar SEDs.

\begin{enumerate}

\item A typical quasar SED from \citet{elv94}. This is a mean SED
  determined from $29$ moderately luminous quasars.

\item A broken power-law SED from \citet{mat87}. This is a mean
  continuum determined by combining direct observations, without
  distinguishing between radio quiet and radio loud quasars.

\item A multi-component continuum with either a small, a medium, or a
  big blue bump (BB) (see the Cloudy document, {\it Hazy} \S\,6.2, for
  details), parameterized by $kT_{\rm BB}=10\mbox{\ eV}$, $100$ eV and
  $300$ eV, respectively \citep{cas06}. By adjusting the parameters
  of this prescription we produce three different strengths of the BB,
  which we list in Table~\ref{tab-specindex} (``small",``medium", and
  ``big"). For each BB strength we examine three different values
  of $\alpha_{\rm ox}$ (denoted as {\sc i}, {\sc ii}, and {\sc iii} in
  Table~\ref{tab-specindex}), which gives us nine different SED shapes
  for this model.

\end{enumerate}

\begin{deluxetable}{lccc}
\tablecolumns{5} 
\tablewidth{0pc}
\tablecaption{Spectral Indices of Different SED Models Compared to Observed Values \label{tab-specindex}}
\tablehead{\colhead{SED\tablenotemark{a}} &
\colhead{$\alpha_{\rm o}$} & \colhead{$\alpha_{\rm x}$} &
\colhead{$\alpha_{\rm ox}$}} \startdata 
Observed, HS1700+6416 & $-0.47$ & $-1.00$ & $-1.87$\\ 
                      &         &         &        \\
Elvis 1994            & $-0.91$ & $-0.93$ & $-1.37$\\
Mathews 1987          & $-0.50$ & $-0.70$ & $-1.41$\\
Small BB {\sc i}      & $-1.24$ & $-0.99$ & $-1.40$\\
Small BB {\sc ii}     & $-1.27$ & $-0.99$ & $-1.60$\\
Small BB {\sc iii}    & $-1.27$ & $-0.99$ & $-1.90$\\
Medium BB {\sc i}     & $-0.54$ & $-1.00$ & $-1.40$\\
Medium BB {\sc ii}    & $-0.56$ & $-1.00$ & $-1.60$\\
Medium BB {\sc iii}   & $-0.56$ & $-1.00$ & $-1.90$\\
Big BB {\sc i}        & $-0.49$ & $-1.82$ & $-1.40$\\
Big BB {\sc ii}       & $-0.50$ & $-2.30$ & $-1.60$\\
Big BB {\sc iii}      & $-0.51$ & $-3.06$ & $-1.90$
\enddata
\tablenotetext{a}{The Roman numerals {\sc i,ii,iii} represent
different values of $\alpha_{\rm ox}$.\smallskip }
\end{deluxetable}

\begin{figure}
\rightline{\includegraphics[width=3.9in]{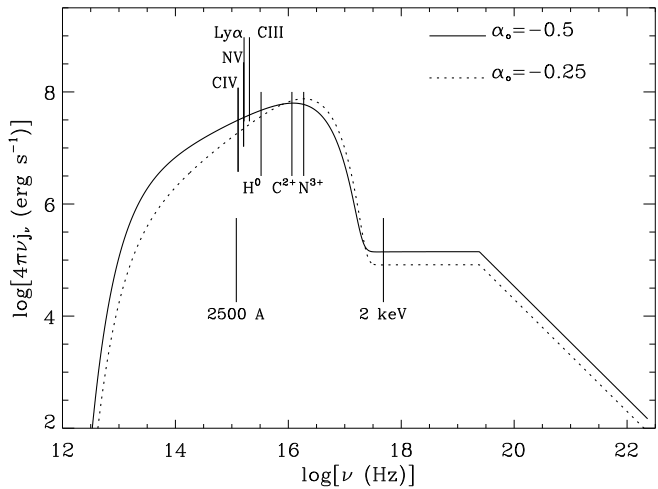}}
\caption{Shapes of two SEDs we used as input
  to our photoionization models. The solid line shows the SED used for
  HS1700+6416 while the dotted line shows the SED used for the other
  two quasars.  These SEDs are normalized to an ionization parameter
  $\log{U}=0.1$ for a hydrogen density of $\log{(n_{\rm H}/{\rm
      cm}^{-3})}=8$. Thus, they have the same photon luminosity
  between 1 and $7.354\times10^{6}$~Ry. Also labelled are positions of
  important transitions -- Ly$\alpha$, \nvdblt\ and \civdblt.  We also
  mark the photon frequencies needed to ionize N$^{+3}$ to N$^{+4}$,
  C$^{+2}$ to C$^{+3}$, and H$^0$ to H$^+$.  We also mark the
  frequencies corresponding to 2500~\AA\ and 2~keV since flux
  densities at these two frequencies are used to define
  \aox. \label{fig-sed}}
\end{figure}

We utilize three parameters to constrain the SED: $\alpha_{\rm
    o}$, $\alpha_{\rm x}$, and $\alpha_{\rm ox}$, which are the
UV/optical spectral slope, X-ray band spectral slope and optical-X-ray
spectral slope, respectively (assuming that $f_\nu\propto
\nu^{+\alpha}$). The last is defined as $\alpha_{\rm
  ox}=0.3838\log\left[\ell_\nu(2\ {\rm
    keV})/\ell_\nu(2500~{\rm\AA})\right]$, where $\ell_\nu(2\ {\rm
  keV})$ and $\ell_\nu(2500~\mbox{\AA})$ are the monochromatic
luminosities at 2~keV and at 2500~\AA\ \citep{tan79}.

The low resolution ($R\sim2000$) UV/optical spectrum of HS1700$+6416$
is available from the Data Release 5 (DR5) quasar catalog
\citep{sch07} of the Sloan Digital Sky Survey (SDSS; York et
al. 2000)\nocite{yor00}.  We fitted the underlying continuum using a
power-law, with a small Balmer bump and the iron emission forest
superimposed \citep{van08}, and obtained $\alpha_{\rm o}=-0.47$ (see
Table~\ref{tab-specindex}).  This object was also observed by {\it
  Chandra} and {\it XMM-Newton}.  The X-ray spectrum was adequately
fit with a power-law index $\sim-1$ ($\alpha_{\rm x}=-1.2\pm0.2$ for
the {\it Chandra} spectrum and $\alpha_{\rm x}=-1.1\pm0.2$ for the
{\it XMM--Newton} spectrum; Misawa et al. 2008)\nocite{mis08}. Thus
we adopted $\alpha_{\rm x}=-1$.

To find $\alpha_{\rm ox}$ we used the empirical correlation between
it and the $2500$~\AA\ quasar luminosity \citep{vig03, str05, ste06}.
We adopt the $\alpha_{\rm ox}$-$\ell_\nu(2500~\mbox{\AA})$ relation by
\citet{ste06}: $$\alpha_{\rm ox}=(-0.137\pm0.008)
\log{\ell_\nu(2500~{\rm\AA})} +\left(2.638\pm0.240\right).$$ The
2500~\AA\ luminosity is calculated as $\ell_\nu(2500~\mbox{\AA})=4\pi
d_{\rm L}^2f_\nu(2500~{\rm\AA})$, in which $d_{\rm L}$ is the
luminosity distance and $f_\nu(2500~{\rm\AA})$ is the rest-frame flux,
derived from an extrapolation of the flux density at
4400\ \AA\ \citep{mis07} assuming $f_\nu\propto\nu^{-0.44}$
\citep{van01}. Thus, for HS1700+6416 we find $\alpha_{\rm ox}=-1.87$

The three spectral indices for HS1700$+6416$, from spectroscopic
observations, are compared to those for possible quasar SEDs
(Table~\ref{tab-specindex}). We find that the Medium BB {\sc iii} SED
best matches the observational constraints thus, we use this SED as
the input ionizing flux for HS1700$+6416$.  This SED
is plotted as a solid line in Fig.~\ref{fig-sed}.

We measured $\alpha_{\rm o}$ of the other two quasars from the
low-resolution spectra of Q1009$+2956$ \citep{bur98} and HE0130$-4021$
\citep{osm94}, respectively.  These UV/optical power-law indices are
determined by connecting two empirical ``line-free'' regions redward
of the Ly$\alpha$ emission line and are $\sim-0.25$ for both
quasars. The 2500~\AA\ luminosities for these two quasars are
determined in the same manner as for HS1700$+6416$ and their
$\alpha_{\rm ox}$ values are both $\sim-1.8$. There have not been any
X-ray spectral observations of Q1009$+2956$ or HE0130$-4021$ so, we
assume they have the same $\alpha_{\rm x}$ as
HS1700$+6416$. Generally, the photoionization cross section decreases
rapidly with photon energy so the ionization state is not sensitive to
the X-ray spectral slope. We then also use the Medium BB {\sc iii}
multi-component SED for both of these quasars, but we adjust the
optical slope to $-0.25$ to match the observations. This SED is shown
as a dotted line in Fig.~\ref{fig-sed}.

We have integrated the above SEDs to compute the bolometric
luminosities and ionizing ($E>1$~Ry) photon rates for the three
quasars ($L_{\rm bol}$ and $Q$, respectively). We first normalized
them to match the 4400~\AA\ luminosity density of each quasar (using
the flux density and luminosity distance from
Table~\ref{tab-sumsys}). The resulting values are listed in
Table~\ref{tab-sumsys}; they will be useful for our estimates of the
properties of the absorbing gas in \S\ref{discussion}.2.

\begin{figure*}
\centerline{
\hfill
\includegraphics[width=3.2in]{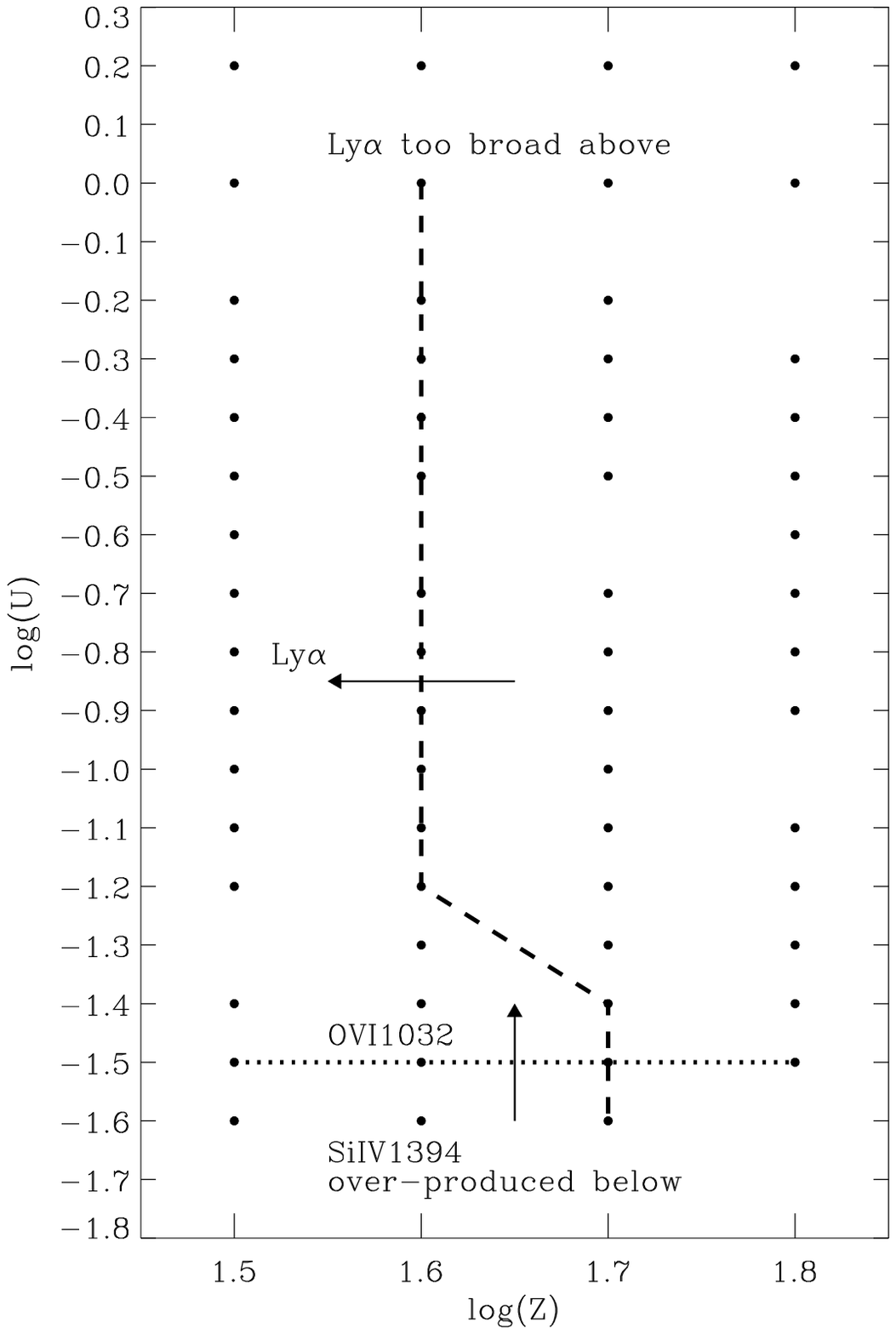} 
\hfill
\includegraphics[width=3.2in]{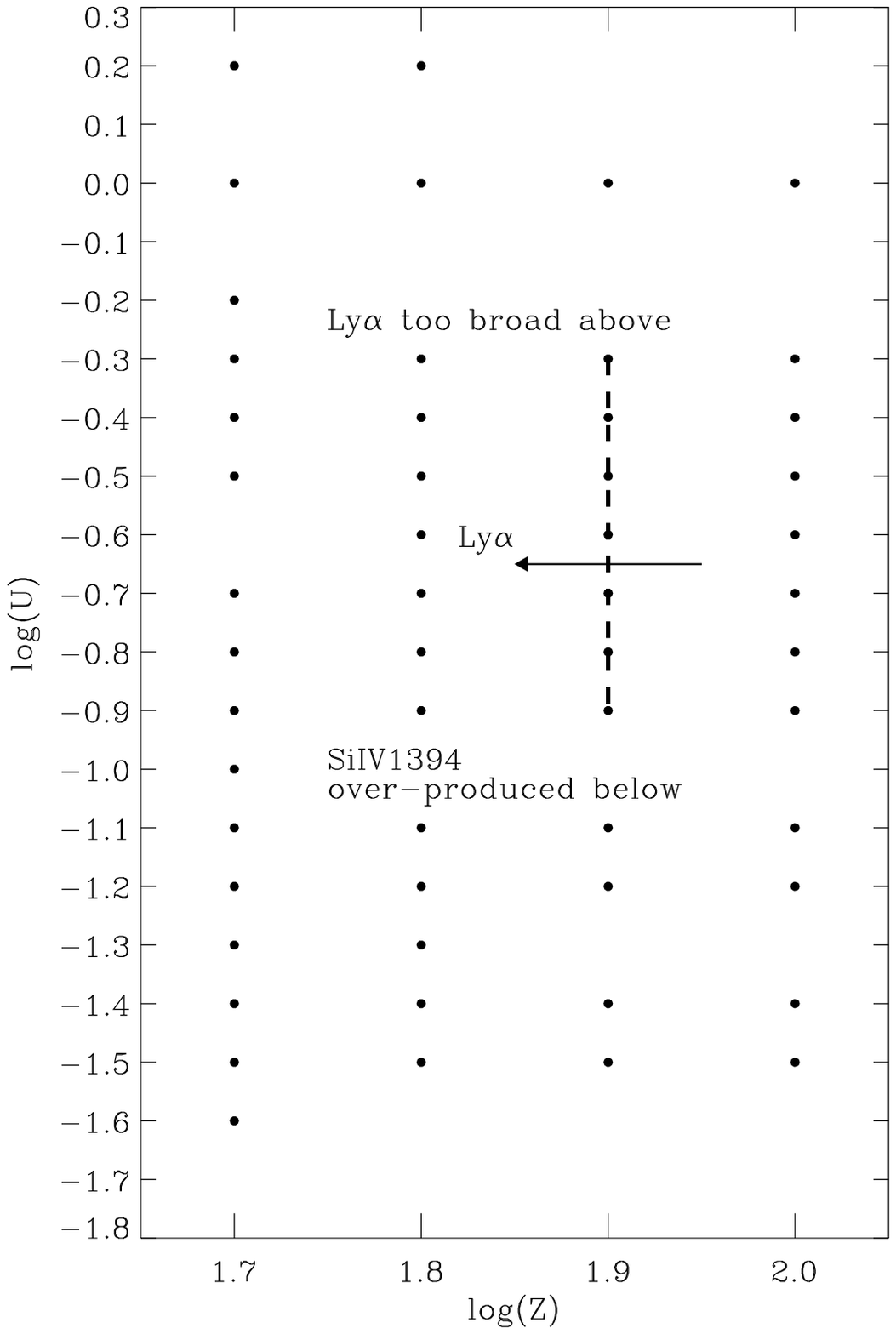} 
\hfill
}
\vspace{1cm} 
\caption{The $\log{(Z/Z_\odot)}$-$\log{U}$ parameter slice at
  $\log{n_{\rm H}}=8$, showing the tracks of acceptable preliminary
  models for the absorption system in the spectrum of
  HE0130$-$4021. The left panel shows the tracks for the blue group
  (kinematic components 1--4, and 9). The right panel shows the tracks
  for the red group (kinematic components 5--8). Black dots are the
  cases we covered in our modelling. The dashed line represents the
  best-fit track for \lya\ while the dotted line represents the
  best-fit track for \ovi. Arrows indicate the directions in which a
  specific theoretical absorption profile becomes stronger, so on the
  ``tail" side, the absorption line is under-produced while on the
  ``head" side, it is over-produced.
\label{fig-q0130-slice}}
\end{figure*}

\begin{deluxetable*}{lccccc}
\tablecolumns{6} 
\tablecaption{Summary of Preliminary Fitting Results \label{tab-summary}}
\tablewidth{0pc} 
\tablehead{ 
\colhead{}                                   &
\colhead{Kinematic}                          &
\colhead{}                                   & 
\colhead{}                                   &
\colhead{}                                   & 
\colhead{Abundance}                         \\
\colhead{Quasar}                             &
\colhead{Component}                          &
\colhead{$\log{(Z/Z_\odot)}$}                & 
\colhead{$\log{U}$}                          &
\colhead{$\log(n_{\rm H}/{\rm cm^{-3}})$}    & 
\colhead{Pattern}                            }
\startdata
HE0130$-$4021\tablenotemark{b} & 1--4,9 & 1.6       & $-0.6$ to 0      & 2      & solar\\
              &        & 1.6--1.7  & $-1.4$ to 0      & 8      & solar\\
              &        & 1.5--1.6  & $-1.4$ to $-0.3$ & 10     & solar\\
              &        & 1.5       & $-1.3$           & 12     & solar\\
              &        & 1.3       & $-1.0$           & 14     & solar\\
\noalign{\vskip 10pt}
              & 5--8   & $\geq1.7$ & $\geq-0.6$       & 2      & solar\\
              &        & 1.9       & $-0.9$ to $-0.2$ & 8      & solar \\
              &        & $\geq1.8$ & $-0.8$ to $0.4$  & 10     & solar\\
              &        & 1.6       & $-0.8$           & 12     & solar\\
              &        & 1.8       & $-0.6$           & 14     & solar\\
\noalign{\vskip 10pt}
Q1009$+$2956  & 1--3   & 2.5       & $0.2$            & 2--8   & solar, [S/H]$\leq1.5$\\
              &        & 2.5       & 0.3              & $9$    & solar, [S/H]$\leq1.5$\\
              &        & 2.4       & 0.2              & 10     & solar, [S/H]$\leq1.4$\\
              &        & 2.3       & 0.3              & 11--12 & solar, [S/H]$\leq1.5$\\
              &        & 2.3       & 0.7              & 12     & solar, [C/H]$\sim2.8$\\
              &        & 2.5       & 0.4              & 13     & solar, [S/H]$\leq1.5$\\
\noalign{\vskip 10pt}
HS1700$+$6416\tablenotemark{a} & 1--2   & 1.4       & 0.5              & 2--8   & solar\\
              &        & 1.3       & 0.4              &  9     & solar\\
              &        & 1.3       & 0.5              & 10     & solar\\
\noalign{\vskip 6pt}
              & 3      & 1.7       & 1.0              & 2--8   & solar
\enddata
\tablecomments{Here we assume that blended kinematic components, as
  indicated in the second column of the table, are described by the
  same combination of model parameters.}
\tablenotetext{a}{The results presented in this table are also a summary of the final
results for this quasar. See the discussion in \S\ref{results}.1 of
the text.
\smallskip}
\end{deluxetable*}

\subsection{Comparison of Cloudy Model Results to the Data and the 
Role of the Coverage Fraction}

Cloudy calculates the electron temperature based upon thermal and
ionization balance.  Using this temperature, we calculate the
microturbulence velocity for N{\,\sc v} based on its observed total
Doppler $b$ parameter assuming that both of them follow Gaussian
distribution.  We then calculate the $b$ parameters for all other
elements by combining this turbulent $b$ with the thermal $b$,
calculated from the atomic weight and the electron temperature.  Using
these $b$ parameters and the column densities of each ion, output by
Cloudy, we synthesize noiseless model spectra, convolving with a
Gaussian function which represents the instrumental profile.  We
compare the theoretical profiles with the observed ones using a
$\chi^2$ test in combination with visual inspection. The wavelength
range within which $\chi^2$ is evaluated is carefully chosen because
even a single pixel affected by a blend or an instrumental a artifact
can dominate the $\chi^2$ values.

We apply a grid method in $(Z/Z_\odot)$--$U$--$n_{\rm H}$ space to
search for acceptable solutions by comparing model profiles to the
observed absorption lines. In order to better visualize the models and
compare them to the observed spectra, we take constant-$n_{\rm H}$
slices and examine combinations of $(Z/Z_\odot)$ and $U$ in this
2-dimensional parameter space.  The ranges of $(Z/Z_\odot)$, $U$, and
$n_{\rm H}$ we explore vary depending on the specific system, but
typically they span $1 \lesssim \log{(Z/Z_\odot)} \lesssim 3$\,\footnote
{In our initial simulations we tried to reproduce the absorption line
  strengths with $\log{(Z/Z_\odot)} \lesssim 1$. However, we found
  that acceptable models require relatively high metalicities, which
  led us to adopt this range of values in our grid search.},
$-1.5\lesssim\log{U}\lesssim1.5$, and $2 \lesssim \log(n_{\rm H}/{\rm
  cm}^{-3}) \lesssim 14$. The initial increments are
$\Delta\log{(Z/Z_\odot)}=0.1$, $\Delta\log{U}=0.1$, and
$\Delta\log(n_{\rm H}/{\rm cm}^{-3})=2$. Finer grids are applied only
if necessary. We assume that different kinematic (Voigt) components
represent different parcels of gas, which we model separately.
However, if kinematic components are blended together, we start by
assuming that they are all described by the same model parameters.  We
refer to such models as the ``initial'' or ``preliminary'' models.
After finding the best preliminary solution, we vary $(Z/Z_\odot)$ and
$U$ for each kinematic component within $0.2$ dex of their preliminary
values to seek better solutions with smaller $\chi^2$ values.

In the process of producing synthetic line profiles for comparison to
the observed spectra, we make use of the coverage fraction inferred
from the relative strengths of the lines in the \nv\ doublet (see
Table~\ref{tab-nvfit}).  Since $C_{\rm f}$ represents the fraction of
photons from the background source that pass through the absorber, we
effectively dilute all the absorption lines of a given system by a
factor equal to $C_{\rm f}$. We have to assume that the same value of
$C_{\rm f}$ applies to all transitions for lack of additional
information. This need not be the case in general, however;
\citet{mis07} do find cases where different resonance doublets in the
same system yield different values of $C_{\rm f}$, for example.  In
one of our three quasars, HS1700$+$6416, we do have separate
measurements of $C_{\rm f}$ for the \nv\ and \civ\ doublets but they
agree with each other within uncertainties. We return to the issue of
the covering factor in Section~\ref{partcov} where we discuss the
possibility of the absorbers covering different fractions of the
continuum and broad-emission line sources.
\section{Photoionization Model Results} \label{results}

\subsection{HE0130$-$4021}

This absorption complex can be described by nine kinematic components
(Fig.~\ref{fig-q0130-sys} and Table~\ref{tab-nvfit}).  These
components are blended so we model them by first assuming that they
are described by the same model parameters. \siivfull\ is not
detected, which yields a lower limit on $\log{U}$. Although \ovidblt\
and \ciiifull\ do not provide strong constraints, we require that they
are not over-produced by our models. The Ly$\alpha$ line
provides a strong constraint for this system. Because the saturated absorption
troughs around 50~km~s$^{-1}$ and 100~km~s$^{-1}$ are not seen in
Ly$\beta$, they cannot be due to Lyman absorption from this
system. These two absorption features actually divide the whole
Ly$\alpha$ absorption system into a red part (positive velocity) and
a blue part (negative velocity), which we can model separately. The blue
part comprises Components 1--4 and 9 while the red part consists of
Components 5--8. Although this system comprises many kinematics components, 
we can still obtain robust results for the coverage fraction of each 
component. This is because we fit the blue and the red members doublet 
simultaneously so that only features appearing in both of doublets are 
contribute to the determination of $C_{\rm f}$. The resulting relative 
uncertainties (listed in Table~\ref{tab-nvfit}) are typically of order 
20\%, which includes errors in profile fitting as well as errors in 
continuum fitting.

\begin{figure}
\figurenum{7a}
\centering
\epsscale{1.1}
\plotone{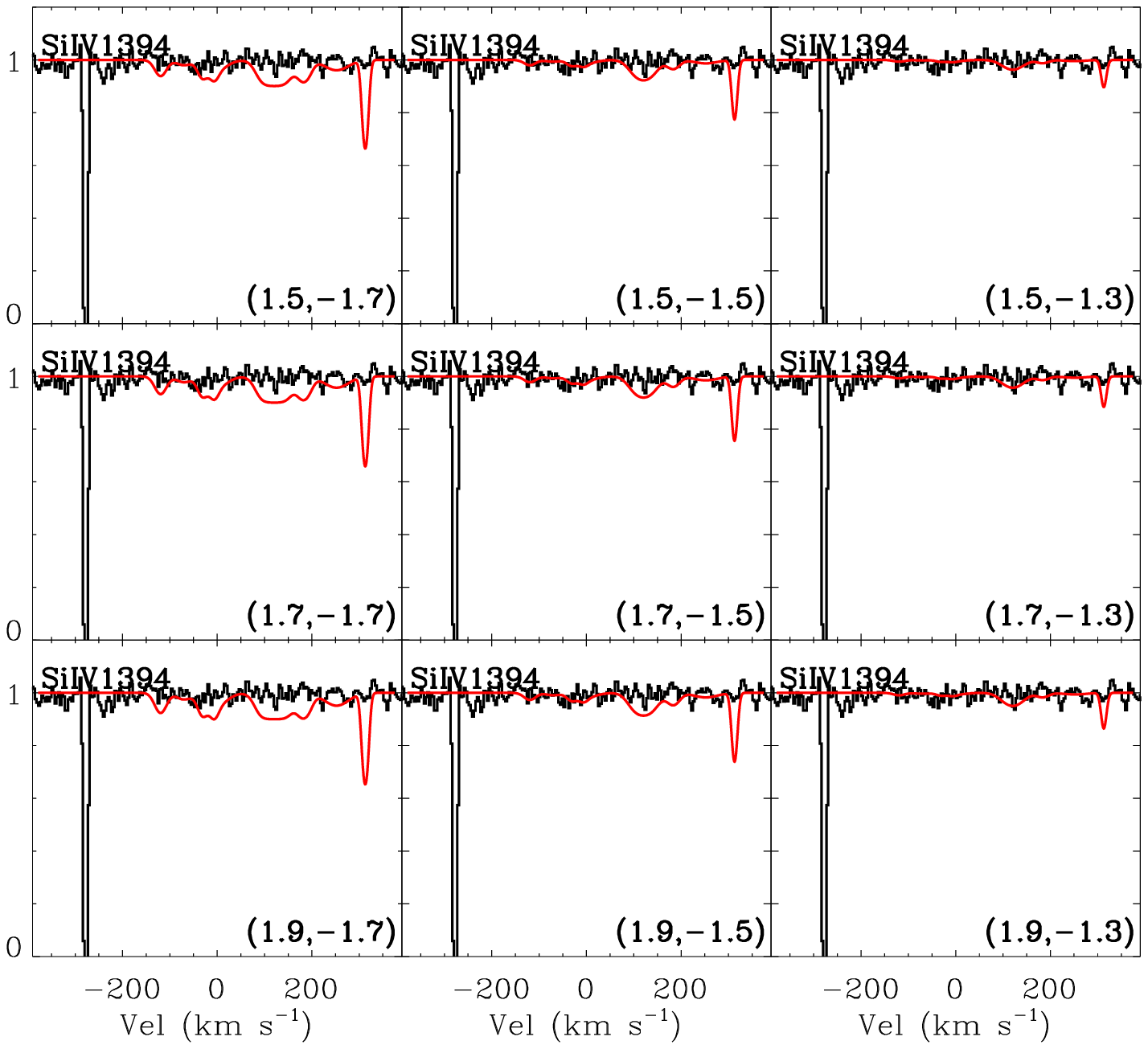}
\vspace{1cm}
\caption{Plots of observed and synthesized profiles of
  all kinematic components of \siivfull\ for the absorption system in
  the spectrum of HE0130$-$4021. We set $\log(n_{\rm H}/{\rm
  cm}^{-3})=8$ and illustrate how the absorption profiles
  change as $(\log{(Z/Z_\odot)},\log{U})$ vary about $(1.7,-1.5)$.
  \label{fig-q0130-n8.c1-9.siiv} }
\vspace{0.5cm}
\figurenum{7b}
\centering
\epsscale{1.1}
\plotone{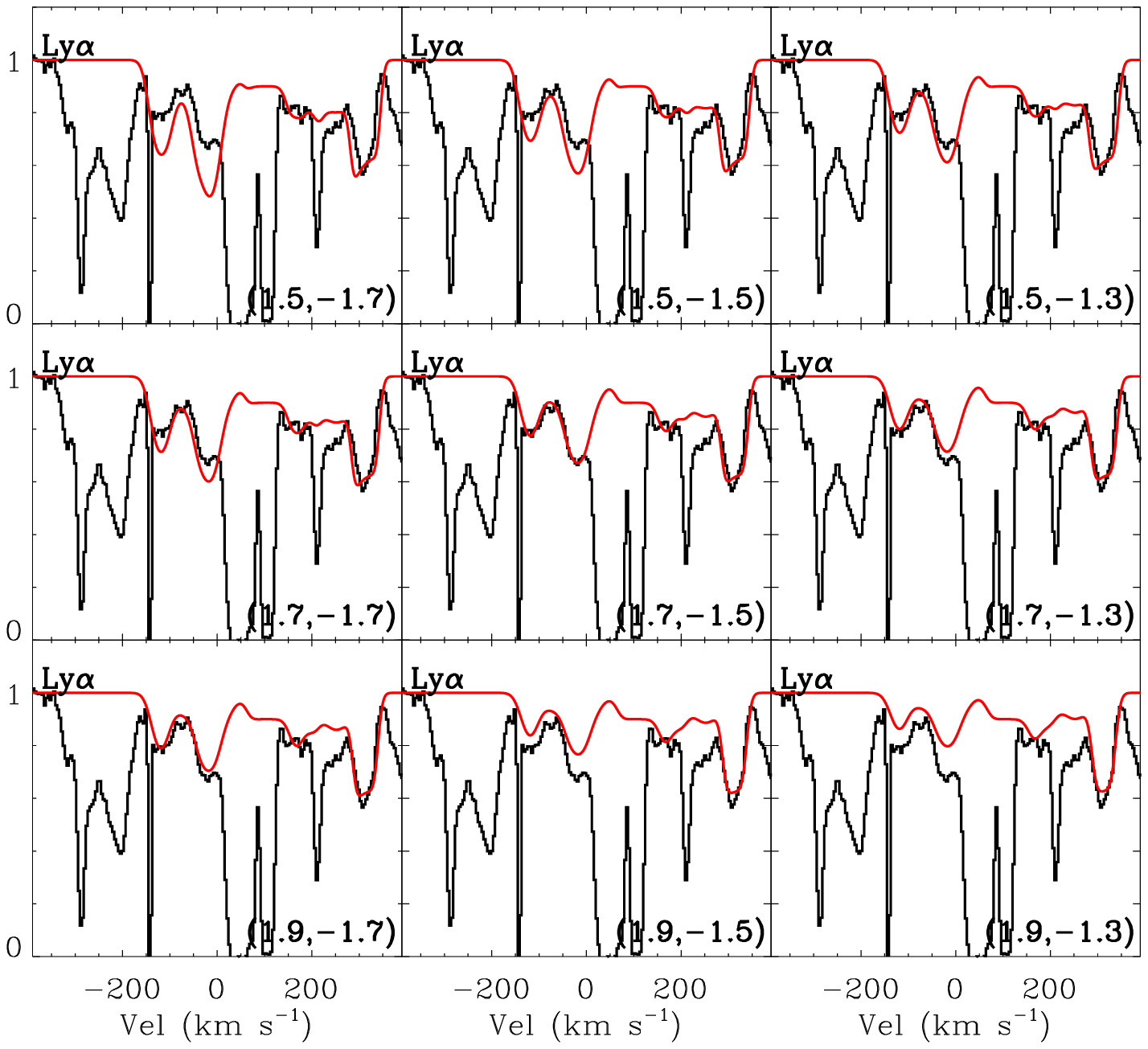}
\vspace{1cm}
\caption{Plots of observed and synthesized profiles of all
  the kinematic components of \lya\ for the absorption system in the
  spectrum of HE0130$-$4021. We set $\log(n_{\rm H}/{\rm
  cm}^{-3})=8$ and illustrate how the absorption profiles
  change as $(\log{(Z/Z_\odot)},\log{U})$ vary about
  $(1.7,-1.5)$.\label{fig-q0130-n8.c1-9.lya}}

\end{figure}

We start by seeking models that can reproduce the observed Ly$\alpha$
absorption complex. For either the red or the blue part of the
profile, we can find a series of solutions for $\log{(Z/Z_\odot)}$ and
$\log{U}$ on a constant-$n_{\rm H}$ slice. Fig.~\ref{fig-q0130-slice}
illustrates the methodology we apply to all systems: it shows the
acceptable models for Ly$\alpha$ for $\log(n_{\rm H}/{\rm cm}^{-3})
=8$ and a range of $\log{(Z/Z_\odot)}$ and $\log{U}$.  The highest and
lowest possible values of $\log{(Z/Z_\odot)}$ and $\log{U}$ for
different values of $\log(n_{\rm H}/{\rm cm}^{-3})$ are tabulated in
Table~\ref{tab-summary}. The blue and red parts of the profile yield
different sets of acceptable model
parameters. Fig.~\ref{fig-q0130-slice} presents constraints on models
based on the observed \ovifull\ and \siivfull\ lines. The effect of
changing $\log{U}$ and $\log{(Z/Z_\odot)}$ on the model profiles of
\siivfull\ and \lya\ is illustrated in
Figures~\ref{fig-q0130-n8.c1-9.siiv}, and
\ref{fig-q0130-n8.c1-9.lya}. The \ovifull\ profile is very insensitive
to metallicity as well as to $\log{U}$ when $\log U \gtrsim-1.5$.
Below this value, the \ovifull\ absorption decreases gradually.
Because the O{\,\sc vi} doublet is in the Ly$\alpha$ forest, we use it
as an upper limit on the O{\,\sc vi} absorption that a model should
produce.  However, at $\log{U}\sim-1.5$, the red parts of O{\,\sc
  vi}~1032,1038 can be adequately fit. Therefore, the intersection
between the acceptable models for the red part of Ly$\alpha$ and the
acceptable models for O{\sc vi}~1032 happens at $\log{U}\sim-1.5$ and
$\log{(Z/Z_\odot)}\sim1.7$.  This model also fits the observed
\ciiifull\ absorption and it does not overproduce \siivfull.  For the
blue part of the system, within the range of models that are
consistent with the Ly$\alpha$ and Si{\sc iv}~1394, the \ovifull\ is
never overproduced.  Since the O{\,\sc vi} is taken as an upper limit,
these models are all considered acceptable, as listed in
Table~\ref{tab-summary}.

The synthesized line profiles are consistent with the data for
values of $\log(n_{\rm H}/{\rm cm}^{-3})$ between 2 and 14.  The
ranges of $\log{(Z/Z_\odot)}$ and $\log{U}$ are slightly different for
different $\log{n_{\rm H}}$ values (Table~\ref{tab-summary}).  An
example of an acceptable model is presented in
Fig.~\ref{fig-q0130-sys}.

There are still discrepancies between the model and the observed
spectrum of \lya. Perhaps the \lya\ line is blended with another
transition because the corresponding absorption trough is not seen in
Ly$\beta$. The $\vel\sim300\mbox{\ km\ s}^{-1}$ absorption trough in the
Ly$\alpha$ window is also underproduced, which seems to suggest an
additional weak component with a small Doppler parameter. Most of the
absorption in O{\,\sc vi} is underproduced by this model, with the
exception of that at $\vel\sim0\mbox{\ km\ s}^{-1}$. The absorption at
this velocity is mainly contributed by Component 9, which is the only
component with a coverage fraction of unity.  We have assumed that the
O{\,\sc vi} absorption is only an upper limit, but if it is real, a
separate phase would be required that has a larger coverage fraction.
In the other two systems O{\,\sc vi} is similarly underproduced; we
discuss this further in Section~\ref{ovi}.

We have attempted to optimise the model parameters for each kinematic
component separately but we found that this exercise did not yield
useful results. We found a large number of ``pockets'' in paramneter
space, spanning a wide range of parameter values, where an acceptable
fit can be achieved but there is no strong reason to prefer one
particular ``pocket'' over the others. We attribute this outcome to
the fact that the Lyman series lines that we use as constraints are
blended with other lines and the profile of \ciii\ has a relatively
low signal-to-noise ratio. Therefore, Table~\ref{tab-summary} provides
the final summary of acceptable model parameters.

\subsection{Q1009$+$2956}

\begin{deluxetable}{cccccc}
\tablecolumns{5} 
\tablecaption{Best model parameters for the Q1009+2956 absorption systems \label{tab-bestq1009}}
\tablewidth{0pc}
\tablehead{
\colhead{Kinem.}            & 
\colhead{log}                     &
\colhead{}                     &
\colhead{log}                     &
\colhead{Abundance}           &
\colhead{$r$ \tablenotemark{a}}\\
\colhead{Comp.}            & 
\colhead{$(Z/Z_\odot)$}  &
\colhead{log$\;{U}$}            &
\colhead{$(n_{\rm H}/{\rm cm^{-3}})$}     &
\colhead{Pattern}              &
\colhead{(pc)}                 }
\startdata
 1 & 2.4 & 0.0 & 2--8 &  solar, [S/H]=1.3   & 10--\\
 2 & 2.6 & 0.3 &      &  solar, [S/H]=1.5   & ~~10,000\\
 3 & 2.3 & 0.0 &      &  solar, [S/H]=1.4   & \\ 
\noalign{\vskip 10pt}
 1 & 2.3 & 0.1 & 9    &  solar, [S/H]=1.3   & $\sim 0.2$\\
 2 & 2.5 & 0.3 &      &  solar, [S/H]=1.5   & \\
 3 & 2.7 & 0.1 &      &  solar, [S/H]=1.7   & \\ 
\noalign{\vskip 10pt}
 1 & 2.4 & 0.0 & 10   &  solar, [S/H]=1.4   & $\sim 0.8$ \\
 2 & 2.5 & 0.3 &      &  solar, [S/H]=1.5   & \\
 3 & 2.3 & 0.0 &      &  solar, [S/H]=1.3   & \\ 
\noalign{\vskip 10pt}
 1 & 2.2 & 0.2 &11--12&  solar, [S/H]=1.2   &  0.04--\\
 2 & 2.4 & 0.3 &      &  solar, [S/H]=1.4   & ~0.2 \\
 3 & 2.4 & 0.5 &      &  solar, [S/H]=1.4   & \\ 
\noalign{\vskip 10pt}
 1 & 2.2 & 0.5 & 12   &  solar, [C/H]=2.7   & $\sim 0.04$  \\
 2 & 2.4 & 0.8 &      &  solar, [C/H]=2.9   & \\
 3 & 2.5 & 0.5 &      &  solar, [C/H]=3.0   & \\ 
\noalign{\vskip 10pt}
 1 & 2.7 & 0.3 & 13   &  solar, [S/H]=1.7   & $\sim 0.004$ \\
 2 & 2.6 & 0.5 &      &  solar, [S/H]=1.6   & \\
 3 & 2.6 & 0.4 &      &  solar, [S/H]=1.6   & 
\enddata
\tablenotetext{a}{The distance of the absorbing gas from the ionizing source
                 for this combination of $U$ and $n_{\rm H}$.}
\end{deluxetable}
 
\begin{figure}
\figurenum{7c}
\centering
\epsscale{1.1}
\plotone{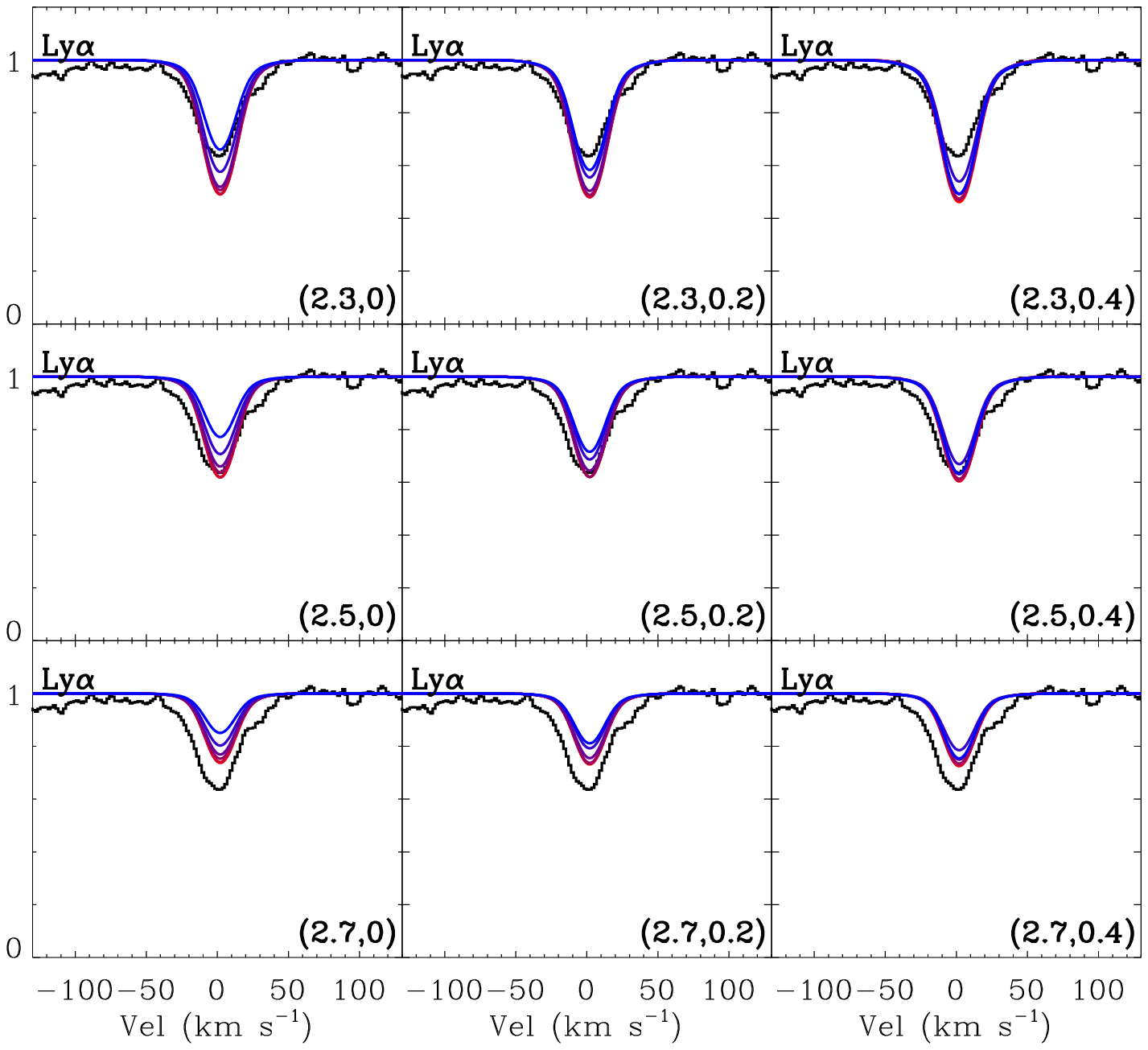}
\vspace{1cm}
\caption{Plots of observed and synthesized profiles of all the
  kinematic components of \lya\ for the absorption system in the
  spectrum of Q1009$+$2956. The different panels show how the models
  change as $(\log{(Z/Z_\odot)},\log{U})$ vary about (2.5,0.2).
  In each panel, the model color changes gradually from red to blue for
 $\log(n_{\rm H}/{\rm cm}^{-3})=2$, 4, 6, 8, 10, and 12.
  \label{fig-q1009-n2-13.c1-3.lya}}
\vspace{0.5cm}
\figurenum{7d}
\centering
\epsscale{1.1}
\plotone{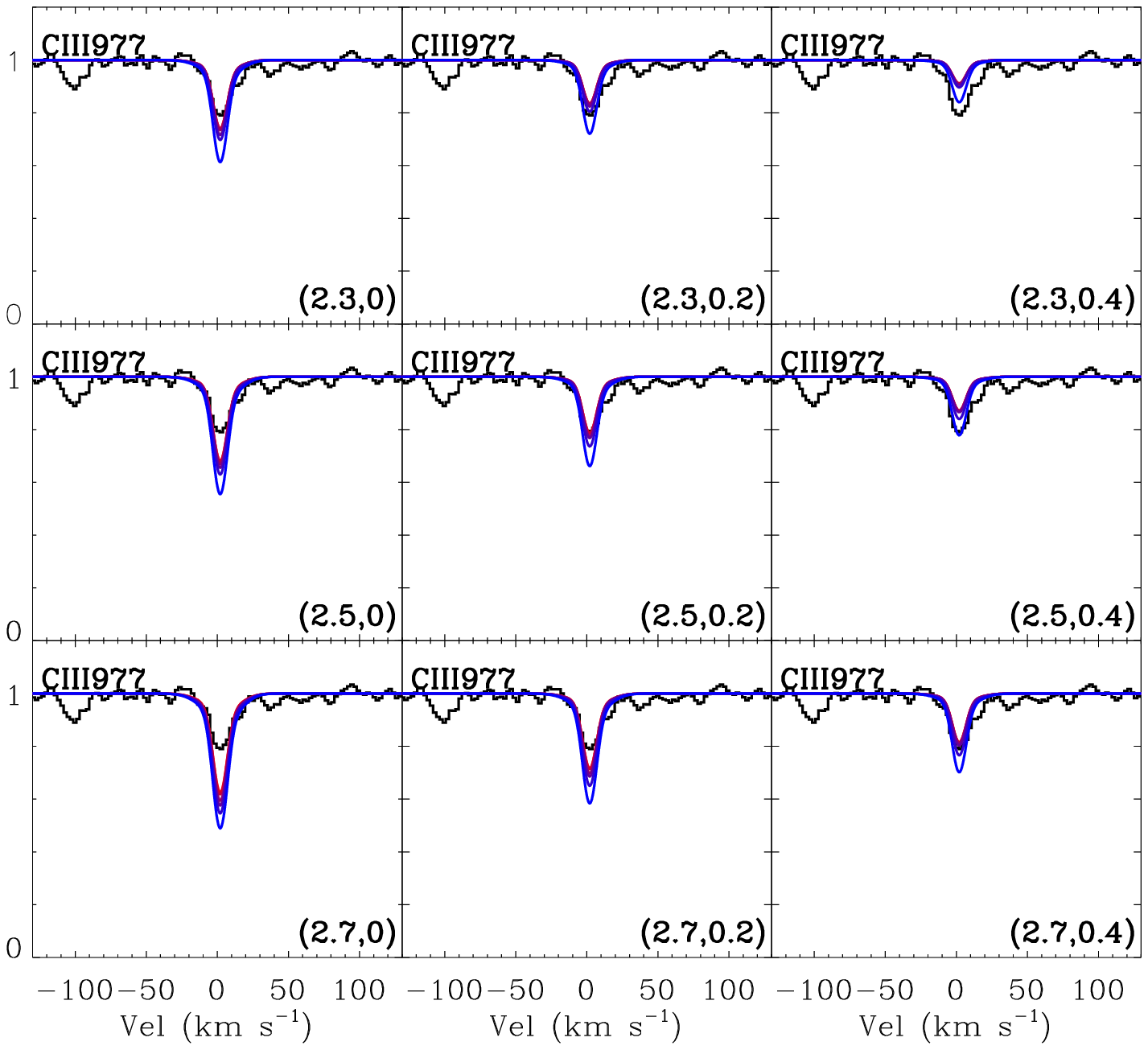}
\vspace{1cm}
\caption{Plots of observed and synthesized profiles of all the
  kinematic components of \ciiifull\ for the absorption system in the
  spectrum of Q1009$+$2956. The different panels show how the models
  change as $(\log{(Z/Z_\odot)},\log{U})$ vary about (2.5,0.2).
  In each panel, the profiles are color-coded according to density in
  the same way as Fig.~\ref{fig-q1009-n2-13.c1-3.lya}.
\label{fig-q1009-n2-8.c1-3.ciii}}
\end{figure}

The absorption-line profile of this system can be reproduced using 3
kinematic components (Fig.~\ref{fig-q1009-sys} and
Table~\ref{tab-nvfit}), which are blended together. In addition to
N{\,\sc v}, there are three main constraints: Ly$\alpha$, \ciiifull,
and \svifull. We initially assume that the three components have the
same parameters. We find that using a solar abundance pattern, we can
always find overlapping regions in parameter space between the best
fits of Ly$\alpha$ and \ciiifull\ if $\log(n_{\rm H}/{\rm cm}^{-3})
\lesssim 13$. These solutions are at
$2.3\lesssim\log{(Z/Z_\odot)}\lesssim2.5$ and
$0.2\lesssim\log{U}\lesssim0.4$,
(Figs.~\ref{fig-q1009-n2-13.c1-3.lya}, and
\ref{fig-q1009-n2-8.c1-3.ciii}). However, we cannot find any
overlapping parameter combinations between suitable fits for \svifull\
and \ciiifull.  The \svifull\ solutions also do not intersect with
acceptable Ly$\alpha$ solutions, except when $\log(n_{\rm H}/{\rm
  cm}^{-3}) \approx 12$. For most densities, only if we lower the
sulphur abundance by about 1 dex (relative to the solar abundance
pattern) can we obtain acceptable model profiles for all three of the
constraints (Table~\ref{tab-summary}). In the case that the absorption
at the position of \svifull\ is contaminated by blends with unrelated
lines, an even larger sulphur abundance reduction would be
required. In principle, it is possible to produce less \svifull\ by
increasing the nitrogen abundance, since \nv\ is the optimized
transition, however we find that such an adjustment is not sufficient
to fully resolve the discrepancy. For the special case of $\log(n_{\rm
  H}/{\rm cm}^{-3}) \approx 12$, the \svifull\ and Ly$\alpha$ profiles
can both be reproduced for $\log{(Z/Z_\odot)} \sim 2.3$ and
$\log{U}\sim0.7$.  For that case, with a higher ionization parameter,
an alternative acceptable model is found if we raise the carbon
abundance by 0.5 dex, leaving the sulfur abundance unchanged from the
solar pattern. These results are summarized in
Table~\ref{tab-summary}.

We next tune separately the metallicity and ionization parameter of
each component, assuming that all three components have the same
volume density, but allowing this common value of the density to
vary. We select the best model parameters at each volume density value
using $\chi^2$ technique and visual inspection
(Table~\ref{tab-bestq1009}). An example of acceptable models is
presented in Fig.~\ref{fig-q1009-sys}.  To obtain this improved fit we
vary the metallicities and ionization parameters of different
components over a range about 0.2 dex.  Through this exercise, we
require that \svifull\ is not overproduced. The left wing of
Ly$\alpha$ cannot be well modelled and probably requires another
component which may not be detected in metal-line absorption or may be
related to a region to the blue of the \nvfull\ profile. We had
neglected that in our fit, since we could not consider whether it was
consistent with N{\,\sc v}~$\lambda1243$. Nevertheless, the modelling
results for this system indicate metallicities considerably higher
than the solar values.

\subsection{HS1700$+$6416}

\begin{deluxetable}{cccccc}
\tablecolumns{5} 
\tablecaption{Best model parameters for the HS1700$+$6416 absorption systems\label{tab-bestq1700}}
\tablewidth{0pc}
\tablehead{
\colhead{Kinem.}            & 
\colhead{log}                     &
\colhead{}                     &
\colhead{log}                     &
\colhead{Abundance}            &
\colhead{$r$ \tablenotemark{a}}\\
\colhead{Comp.}            & 
\colhead{$(Z/Z_\odot)$}  &
\colhead{$\log{U}$}            &
\colhead{$(n_{\rm H}/{\rm cm^{-3}})$} &
\colhead{Pattern}              &
\colhead{(pc)}                 }
\startdata
 1 & 1.3 & 0.5 & 2--8  &  solar & 9--9000\\
 2 & 1.6 & 0.6 &       &  solar & \\
\noalign{\vskip 10pt}
 1 & 1.2 & 0.5 &  10   &  solar & {$\sim 1$}\\
 2 & 1.5 & 0.6 &       &  solar & \\
\noalign{\vskip 10pt}
 3 & 1.7 & 1.0 & 2--8 &  solar & 5--5000
\enddata
\tablenotetext{a}{The distance of the absorbing gas from the ionizing source
                 for this combination of $U$ and $n_{\rm H}$.}
\end{deluxetable}

\begin{figure}
\figurenum{7e}
\centering
\epsscale{1.1}
\plotone{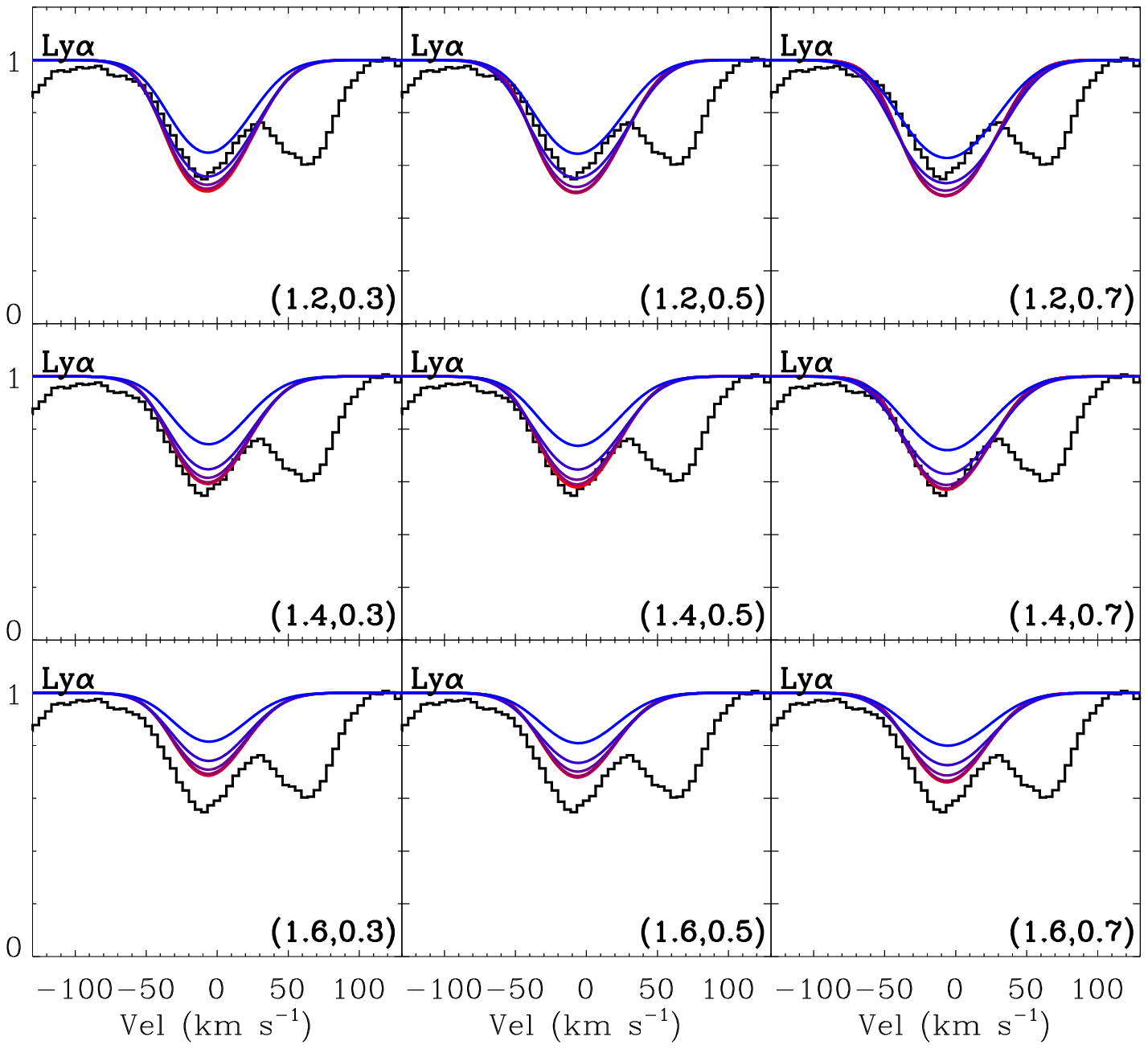}
\vspace{1cm}
\caption{\label{fig-q1700-n2-12.c1-2.lya}Plots of observed and synthesized
 profiles of the two central kinematic components of Ly$\alpha$ for
 the absorption system in the spectrum of HS1700$+$6416. The
 different panels show how the models change as
 $(\log{(Z/Z_\odot)},\log{U})$ vary about (1.4,0.5).  In each panel,
 the profiles are color-coded according to density in the same
 way as Fig.~\ref{fig-q1009-n2-13.c1-3.lya}. 
 Figures~\ref{fig-q0130-n8.c1-9.siiv}--\ref{fig-q1009-n2-8.c1-3.ciii}, 
 \ref{fig-q1700-n2-12.c3.civ}, and \ref{fig-q1700-n2-12.c3.lya} are 
 available in the online version of the Journal.}
\vspace{0.5cm}
\figurenum{7f}
\centering
\epsscale{1.1}
\plotone{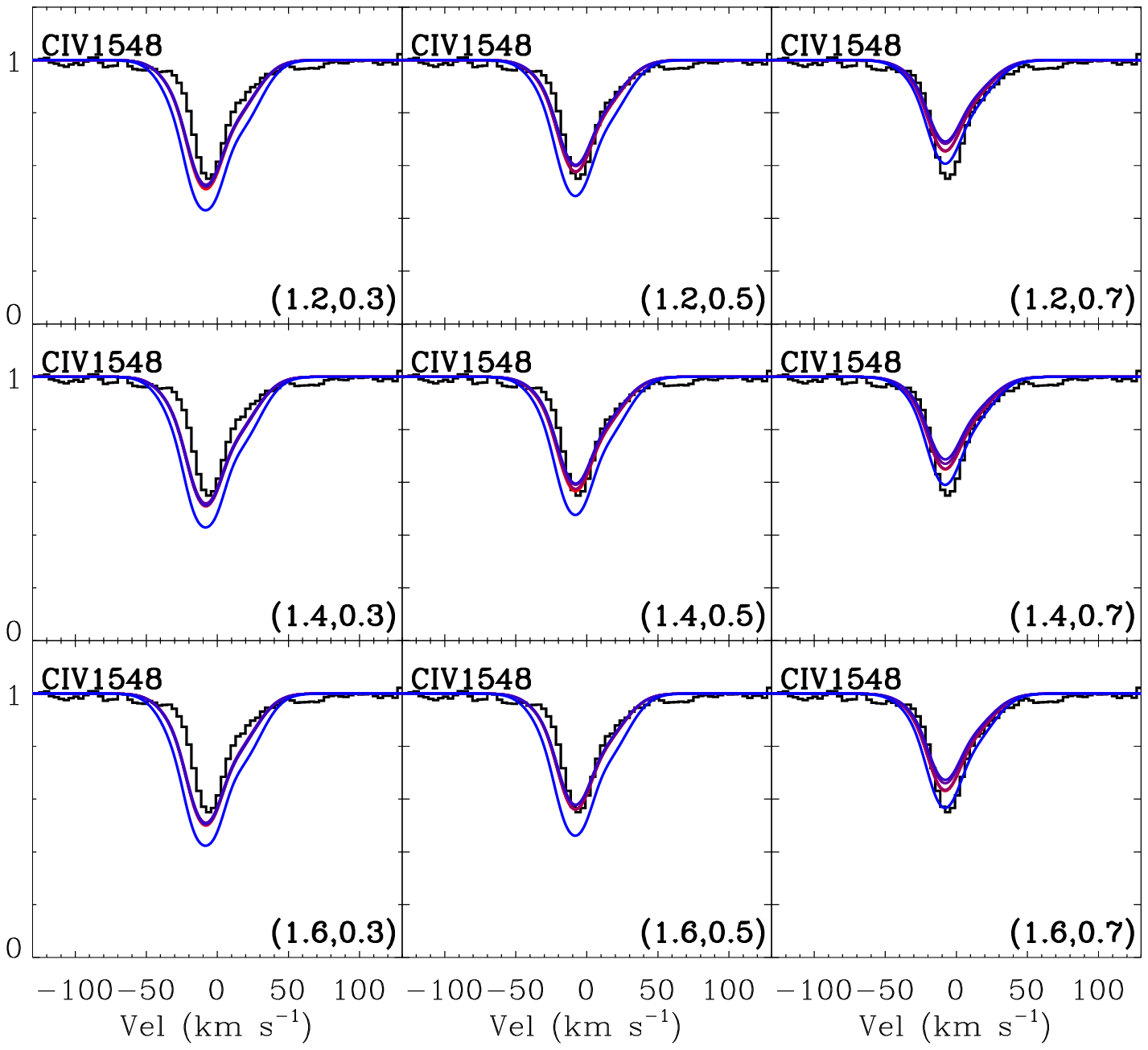}
\vspace{1cm}
\caption{Plots of observed and synthesized profiles of the
  two central kinematic components of \civfull\ for the absorption
  system in the spectrum of HS1700$+$6416. The different panels show
  how the models change as $(\log{(Z/Z_\odot)},\log{U})$ vary
  about (1.4,0.5).  In each panel, the profiles are color-coded
  according to density in the same way as
  Fig.~\ref{fig-q1009-n2-13.c1-3.lya}.
  \label{fig-q1700-n2-12.c1-2.civ}}
\label{fig:fig1b}
\end{figure}

This absorption system can be described by $3$ kinematic components
(Fig.~\ref{fig-q1700-sys}, Table~\ref{tab-nvfit}). Because the third
component, at $\vel\sim315\mbox{\ km\ s}^{-1}$, is well separated from
the central two components, we determine its parameters
independently. Assuming the central two clouds to have the same
parameters, the preliminary best fits, consistent with both Ly$\alpha$
and \civfull\, as well as with undetected transitions, are tabulated
in Table~\ref{tab-summary}. In Figures~\ref{fig-q1700-n2-12.c1-2.lya}
and \ref{fig-q1700-n2-12.c1-2.civ} we plot the profiles of the two
central components of Ly$\alpha$ and C{\,\sc iv} in the
$\log{(Z/Z_\odot)}$--$\log{U}$ parameter slice around this preliminary
best fit for $\log(n_{\rm H}/{\rm cm}^{-3})=2$, 4, 6, 8, 10, and
12. For $2 \lesssim \log(n_{\rm H}/{\rm cm}^{-3}) \lesssim8$, the line
strengths are almost independent of density, but for higher densities
we see weaker \lya\ and slightly stronger \civ\ absorption. When
$\log(n_{\rm H}/{\rm cm}^{-3}) \gtrsim 11$, the acceptable fits to
\civfull\ occur at $\log{U}\sim0.6$, but Ly$\alpha$ is underproduced
by the models so that it is impossible to obtain an acceptable fit at
these very high densities. Observationally, variability on a timescale
of 6.5 months in the quasar's rest-frame implies $\log(n_{\rm H}/{\rm
  cm}^{-3})>3~$ (based on considerations of the recombination time;
see \S\ref{data}).

To refine our preliminary models, we next allow the central two
components to have different parameters.  Thus, we adjust
$\log{(Z/Z_\odot)}$ and $\log{U}$ of each cloud separately, within
$\pm0.2$ dex of the preliminary best fit. A complete tabulation of the
best model parameters at each value of the volume density can be found
in Table~\ref{tab-bestq1700}. Take $\log (n_{\rm H}/{\rm cm}^{-3})=8$
for example, a better fit is found at $\log{(Z/Z_\odot)}=1.3$,
$\log{U}=0.5$ for the first component and $\log{(Z/Z_\odot)}=1.6$,
$\log{U}=0.6$ for the second component.  Even for this example, shown
in Fig.~\ref{fig-q1700-sys}, the \lya\ components at $\vel
\sim0\mbox{\ km\ s}^{-1}$ are slightly underproduced.  This may imply
that an additional component would be needed, or that the \lya\ is
blended with a line from a system at another redshift.  The latter is
not unlikely given the unidentified component present at $\vel
\sim70\mbox{\ km\ s}^{-1}$

\begin{figure}
\figurenum{7g}
\centering
\epsscale{1.1}
\plotone{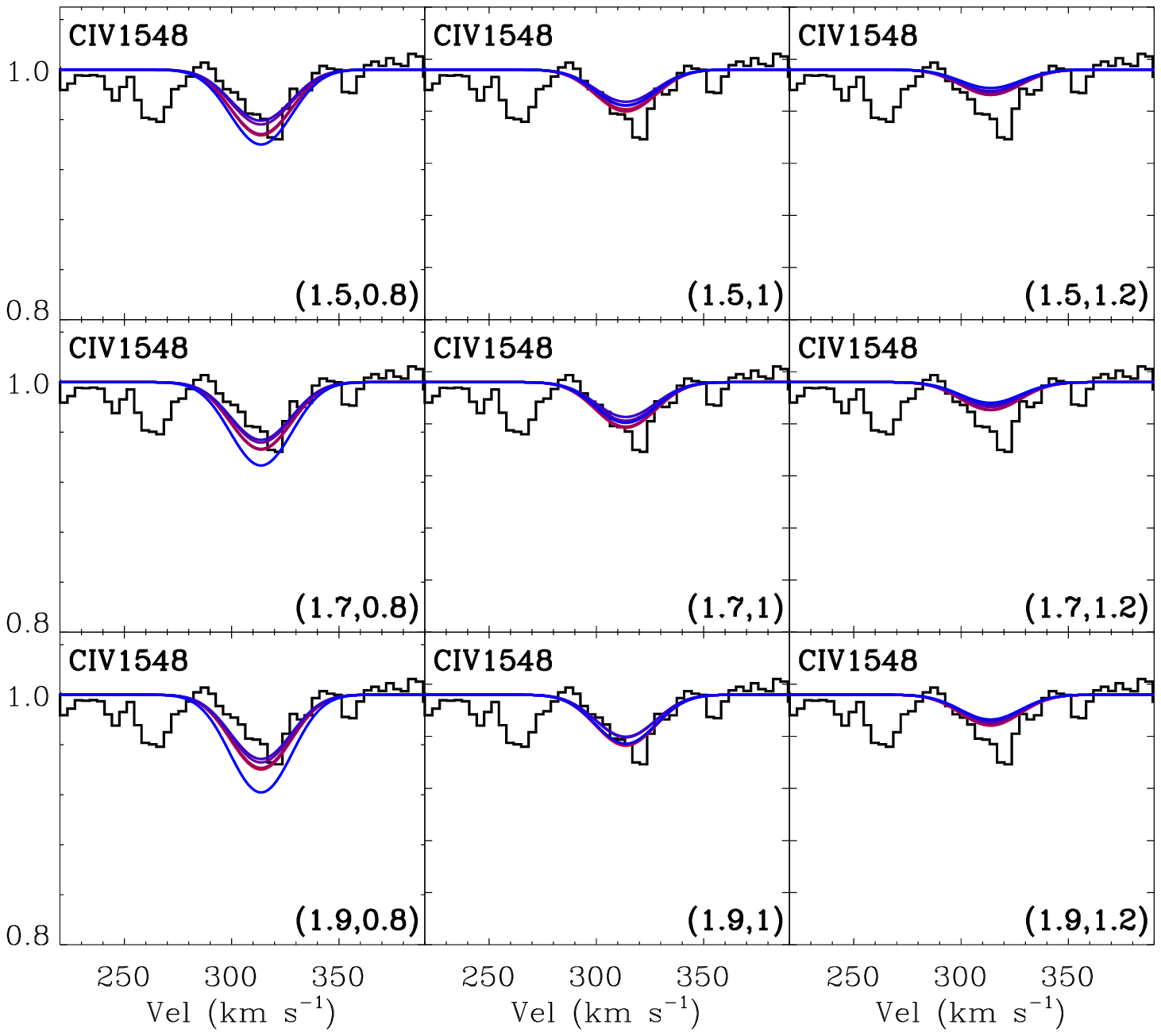}
\vspace{1cm}
\caption{Plots of observed and synthesized profiles of the kinematic
  component at 315~\kms\ of \civfull\ for the absorption system in the
  spectrum of HS1700$+$6416. The different panels show how the models
  change as $(\log{(Z/Z_\odot)},\log{U})$ vary about (1.7,1.0).
  In each panel, the profiles are color-coded according to density in the same
  way as Fig.~\ref{fig-q1009-n2-13.c1-3.lya}.  The vertical axis scale
  is expanded in order to display this weak feature.
  \label{fig-q1700-n2-12.c3.civ}}
\label{fig:fig1c}
\vspace{0.5cm}
\figurenum{7h}
\centering
\epsscale{1.1}
\plotone{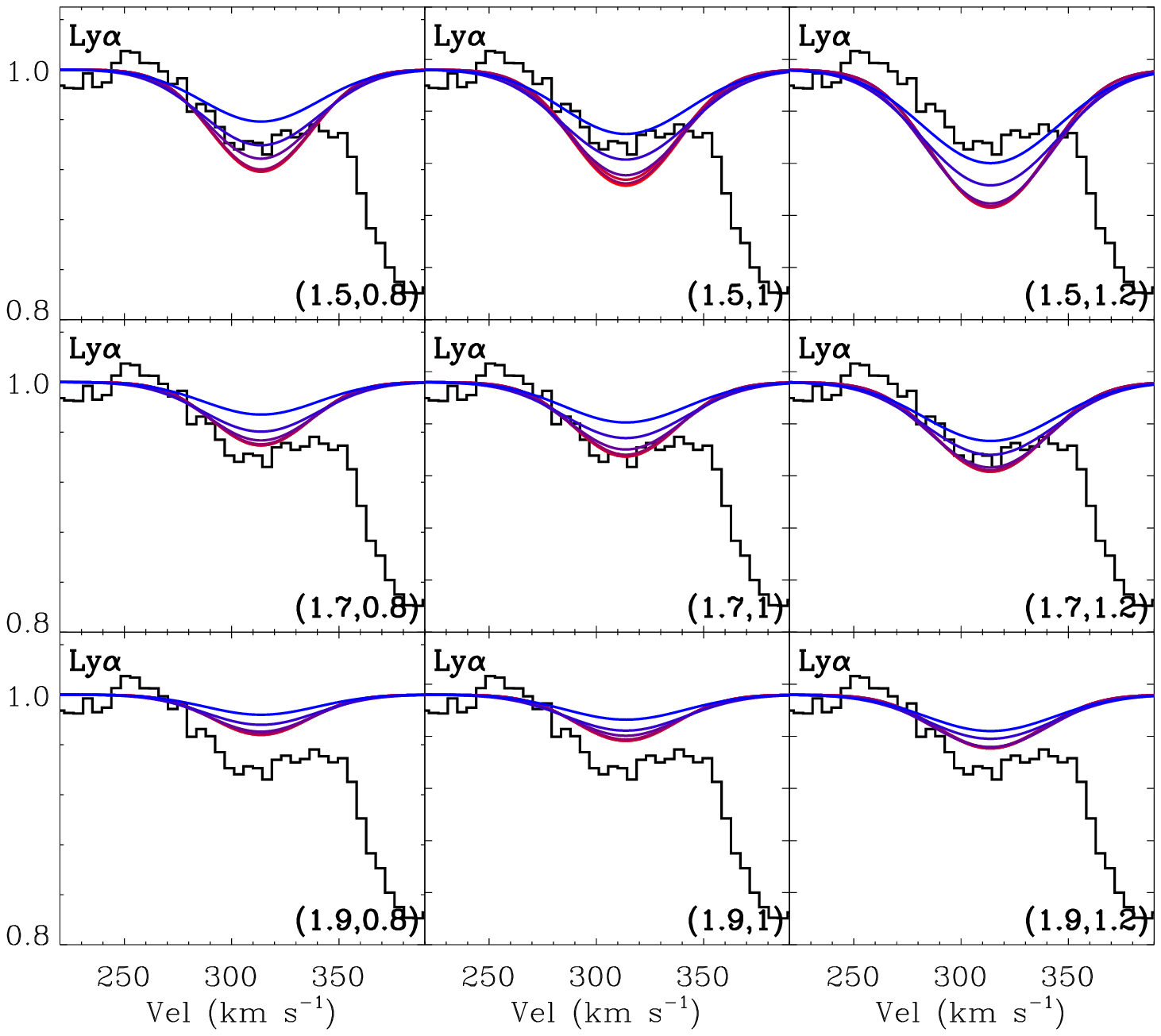}
\vspace{1cm}
\caption{Plots of observed synthesized profiles of the kinematic
    component at 315~\kms\ of \lya\ for the absorption system in the
    spectrum of HS1700$+$6416. The velocity zero point is defined as in
    Fig.~\ref{fig-q0130-sys}. In each panel, the profiles are
    color-coded according to density in the same way as
    Fig.~\ref{fig-q1009-n2-13.c1-3.lya}.  The vertical axis scale is
    expanded in order to display this weak feature.
    \label{fig-q1700-n2-12.c3.lya}}
\end{figure}

The best fit for the third kinematic component at $\vel\sim315\mbox{\ km\
  s}^{-1}$ is achieved at $\log{(Z/Z_\odot)}=1.7$, $\log{U}=1$ at
$\log (n_{\rm H}/{\rm cm}^{-3})=2$--$8$. We cannot obtain any
acceptable fits for $\log(n_{\rm H}/{\rm cm}^{-3}) > 8$ because model
Ly$\alpha$ profiles become broader than the observed profiles for
cases that match the observed C{\,\sc iv}
(Figures.~\ref{fig-q1700-n2-12.c3.civ} and \ref{fig-q1700-n2-12.c3.lya}).
Combining all three components, Fig.~\ref{fig-q1700-sys} shows one of
the acceptable fits we obtain for this system.

\subsection{The O{\,\sc vi} Doublet}\label{ovi}

The energy required to produce \ovi\ is much higher (8.371~Ry) than
\civ\ (3.520~Ry) and \nv\ (5.694~Ry). Thus, the \ovi\ doublet probes
highly-ionized gas and could, in fact, arise in a more tenuous gas
that is not co-spatial with the gas responsible for the
lower-ionization absorption lines and also has a different coverage
fraction of the background source(s). With the above in mind, we note
that in the case of HS1700$+$6416 (Fig.~\ref{fig-q1700-sys}), the
best model cannot fully reproduce the \ovi\ absorption in both the
central and redward components. This could signal that the \ovi\
absorber has a larger coverage fraction.  Another possibility is that
the \ovi\ is contaminated by lines from the Lyman forest, which is
supported by the large widths of the ``black'' troughs around the
O{\,\sc vi~$\lambda1038$} line of HS1700$+$6416 and Q1009$+$2956 (
Figures.~\ref{fig-q1009-sys} and \ref{fig-q1700-sys}).  Almost all
spectral regions around O{\,\sc vi} for the three quasars have some
indication of blending, which makes the comparison of the synthesized
O{\,\sc vi} doublet profile to the data ambiguous. Therefore, we can
only use the observed \ovi\ absorption troughs to set an upper limit
on the absorption produced by our models.
\section{Partial Coverage of Continuum and Broad-Emission Line Regions} \label{partcov}

In principle, the intrinsic absorbers could cover different fractions
of the continuum source and the BELR \citep[e.g.,][]{gan99}. Thus an
absorption line superposed on a broad emission line represents the
absorption of different fractions of emission-line and continuum
photons.  If this is true, the coverage fraction one obtains from
fitting a doublet is an {\it effective} coverage fraction, which
represents the fraction of the photons from all the background sources
that pass through the absorber. Thus, for an absorption line that is
superposed on a broad emission line and the continuum, a more general
expression for the normalized residual flux at a certain velocity is
given by \citep[see][]{gan99}
\begin{equation}
  \label{flux}
  R(\vel)=1-\frac{\left[1-e^{-\tau(\vel)}\right]
  \left[C_{\rm c}(\vel)+W(\vel)\,C_{\rm e}(\vel)\right]}{1+W(\vel)}
\end{equation}
\citep[see also the discussion and more general formalism
in][]{gab05}.  In the above expression, $C_{\rm c}$ and $C_{\rm e}$
represent the coverage fractions of the continuum source and of the
BELR and $W=f_{\rm e}/f_{\rm c}$ is the ratio of the flux contribution
of the emission line (without continuum) and the continuum only at the
position of the absorption line. In principle $C$ and $W$ are
functions of velocity across the absorption-line profile.  However,
since the absorption lines of interest here are fairly narrow, $W$
does not change appreciably over the profile of the absorption line and
we can regard it as independent of velocity. By rearranging equation
(\ref{flux}) one can show that the effective coverage fraction,
$C_{\rm f}$, can be expressed as the weighted average of $C_{\rm e}$
and $C_{\rm c}$ as
\begin{equation}
  \label{cecc}
  C_{\rm f}=\frac{C_{\rm c}+WC_{\rm e}}{1+W}.
\end{equation}
In practice, we can determine $C_{\rm f}$ by fitting the profiles of
the UV resonance doublets and $W$ by measuring the strength of the
broad emission line relative to the continuum at the location of the
absorption doublet. This leads to a relation between $C_{\rm e}$ and
$C_{\rm c}$ but it does not allow us to determine $C_{\rm e}$ and
$C_{\rm c}$ separately. An example of a graphical depiction of this
relation can be found in Fig.~10 of \citet{gan99}.

The \nv\ absorption doublet presents a further complication because it
is likely to be superposed on the {\it blend} of the \lya\ and \nv\
emission lines.  The regions emitting \lya\ and \nv\ are likely to
have a different spatial extent and their coverage fractions should
also be different. This is suggested by the results of reverberation
mapping studies \citep[see, for example][]{peterson00, onken02,
kollatschny03}, which indicate an ionization stratification in the
BELR. Under these circumstances, in the numerator of
equations~(\ref{flux}) and (\ref{cecc}), we must replace $WC_{\rm e}$
by $W_{\rm Ly\alpha} C_{\rm Ly\alpha}+W_{\rm N\,V} C_{\rm N\,V}$ and
in the denominator, we must replace $W$ by $W_{\rm Ly\alpha} + W_{\rm
N\,V}$. In this new notation $C_{\rm Ly\alpha}$ and $C_{\rm N\,V}$ are
the coverage fractions of the \lya\ and \nv\ emitting regions and
$W_{\rm Ly\alpha}$ and $W_{\rm N\,V}$ are defined in a manner
analogous to the definition of $W$.

Putting aside the above complication, we can invert
equation~(\ref{cecc}) to obtain an expression for $C_{\rm c}$, in
terms of $C_{\rm f}$, $W$, and $C_{\rm e}$. However, since $C_{\rm e}$
is unknown in practice, this relation can only lead to upper and lower
bounds on $C_{\rm c}$, corresponding to $C_{\rm e}=0$ and $1$,
respectively. Even if the absorption doublet is superposed on a blend
of emission lines, the expression for $C_{\rm f}$ reduces to
equation~(\ref{cecc}) if the BELR is fully
covered or not covered at all. Taking the error bars on the measured
quantities into account we obtain the following expressions for the
limits on $C_{\rm c}$:
 
\begin{eqnarray}
C_{\rm c}^{\rm max} & = & \min\left\{1,\;
[1+(W+\delta_W)](C_{\rm f}+\delta_{C_{\rm f}})\right\} 
\; {\rm and} \label{ccmax} \\
C_{\rm c}^{\rm min} & = & \max\left\{0,\;
[1+(W-\delta_W)](C_{\rm f}-\delta_{C_{\rm f}})-(W+\delta_W)\right\}\; ,
\label{ccmin}
\nonumber
\end{eqnarray}
where $\delta_X$ denotes the error bar on $X$.
 
In an analogous manner, we can use equation~(\ref{cecc}) to derive
limits on $C_{\rm e}$, using the fact that $C_{\rm c}$ must lie between
0 and 1. The corresponding expressions, including uncertainties on the 
measured quantities, are:
 
\begin{eqnarray}
C_{\rm e}^{\rm max} & = & \min\left\{1,\;
{[1+(W+\delta_W)](C_{\rm f}+\delta_{C_{\rm f}})\over (W-\delta_W)}\right\} 
\; {\rm and} \label{cemax}\\
C_{\rm e}^{\rm min} & = & \max\left\{0,\;
{[1+(W-\delta_W)](C_{\rm f}-\delta_{C_{\rm f}})\over (W+\delta_W)-1}\right\}\; .
\label{cemin}\\
\nonumber
\end{eqnarray}

We have used flux-calibrated, low-resolution spectra of the three
quasars to measure the values of $W$ for the \lya, and \nv\ absorption
lines (as well as \civ\ absorption line in the case of HS1700$+$6416).
Using these values of $W$ and the values of $C_{\rm f}$ from 
Table~\ref{tab-nvfit} we have estimated the
limits on the coverage fractions of the continuum source and the BELR,
$C_{\rm c}$ and $C_{\rm e}$, which we list in Table~\ref{tab-cecc}.
Limits derived from the \lya\ absorption line involve the assumption
that the effective coverage fraction of the \lya\ absorber is the
average of the values measured for the \nv\ absorber, which need not
be correct. Nevertheless, these limits appear to be consistent with
those derived from the \nv\ and \civ\ doublets. The limits on the
continuum coverage fraction derived from different \nv\ kinematic
components in the same quasar appear to be consistent with each other
with one exception: component 9 of HE0130$-$4021, for which the
effective coverage fraction is measured to be $C_{\rm f}=1$.  The only
reliable limit on the BELR coverage fraction comes from the \civ\
doublet of HS1700+6416.

\begin{deluxetable}{llccc}
\tablecolumns{5} 
\tablewidth{0pc}
\tablecaption{Constraints on Continuum and BELR Coverage Fractions \label{tab-cecc}} 
\tablehead{
\colhead{Quasar} &
\colhead{Ion} &
\colhead{$W\;$\tablenotemark{a}} & 
\colhead{$C_{\rm c}$} &
\colhead{$C_{\rm e}$}
}
\startdata
HE0130$-$4021 & N{\,\sc v}   & $1.1\pm0.4$             & unconstr.     & \dots \\
              &              &                         & $< 0.5$       & \dots \\
              &              &                         & unconstr. & \dots \\
              &              &                         & $< 0.4$       & \dots \\
              &              &                         & $< 0.3$       & \dots \\
              &              &                         & $< 0.4$       & \dots \\
              &              &                         & unconstr. & \dots \\
              &              &                         & $< 0.6$       & \dots \\
              &              &                         & $> 0.3$       & \dots \\
              & Ly$\alpha$   & $0.5\pm0.3$             & unconstr. & unconstr.\\
              & ($C_{\rm f}=0.3$)\,\tablenotemark{b} & &           & \\
\noalign{\vskip 10pt}
Q1009$+$2956  & N{\,\sc v}  & $0.66\pm0.12$           & unconstr. & \dots \\
              &              &                         & $> 0.2$       & \dots \\
              &              &                         & $< 0.5$       & \dots \\
             & Ly$\alpha$   & $1.2\pm0.2$   & $< 0.6$       & unconstr.\\
              & ($C_{\rm f}=0.5$)\,\tablenotemark{b} & &           & \\
\noalign{\vskip 10pt}
HS1700$+$6416 & N{\,\sc v}   & $0.96\pm0.06$           & 0.2--0.8      & \dots \\
              &              &                         & $< 0.7$       & \dots \\
              &              &                         & $< 0.8$       & \dots \\
              & C{\,\sc iv}  & $0.41\pm0.02$           & 0.2--0.8      & $> 0.4$\\
              & Ly$\alpha$   & $0.6\pm0.2$   & $< 0.7$       & $> 0.2$\\
              & ($C_{\rm f}=0.4$)\,\tablenotemark{b} & &           & \\
\enddata
\tablenotetext{a}{The values of $W$ were obtained from the following
spectra: for HE0130$-$4021 we used the spectrum published by
\citet{osm76}, for Q1009$+$2956 we used the spectrum published by
\citet{bur98}, and for HS1700$+$6416 we used the SDSS DR5 spectrum.}
\tablenotetext{b}{In the case of Ly$\alpha$ we cannot determine
$C_{\rm f}$ from the data, therefore we assign a value equal to the
average of all the N{\,\sc v} kinematic components.\smallskip}
\end{deluxetable}

A closely related issue is whether the superposition of absorption
lines on strong emission lines dilutes their strengths by different
amounts and leads to large errors in our estimated abundances. The
relative strengths of the \lya\ and \nv\ absorption lines are
particularly sensitive to this effect, especially when the blueshift
of the absorption lines is small. In a specific scenario where the
absorber covers only the continuum source, the \lya\ absorption line
is superposed on the peak of the \lya\ emission line and can be
diluted much more than the \nv\ line which is superposed on the red
wing of the \lya\ emission line (the velocity difference between the
\lya\ and \nv\ absorption lines is approximately 5925~\kms). In such a
case, the procedure we follow here to determine column densities, will
lead us to underestimate the column density of \lya\ relative to that
of \nv\ and our models would yield a nitrogen abundance that is higher
than the true one.

Examining the values of $W$ listed in Table~\ref{tab-cecc}, we see
that if such a scenario is true, the \nv\ absorption line is actually
diluted more than the \lya\ absorption line in HE0130$-$4021 and
HS1700+6416. In other words, in these two quasars we would have
underestimated the nitrogen abundance rather than overestimated it.
To verify our expectation quantitatively, we explored a new set of
models for HS1700+6416 in which we assumed that the absorber covers
the continuum source but not the BELR. Thus, we started by adopting
the $C_{\rm f}$ value of Cloud 2 because it is the dominant component
and set $C_e=0$ and $C_c=0.64$ for all transitions.  As a result, the
{\it effective} coverage fraction for each transition depends on its
$W$ value, e.g., $C_{\rm f}=0.33$ for N{\sc v}, $C_{\rm f}=0.45$ for
C{\,\sc iv}, and $C_{\rm f}=0.34$ for Ly$\alpha$. By setting
$\log(n_{\rm H}/{\rm cm}^{-3})=8$, we find an acceptable fit at
$(\log{(Z/Z_\odot)},\log{U})=(1.6,0.8)$.  Compared to our best fit
with $C_{\rm c} = C_{\rm e}$, this metallicity is 0.2~dex higher and
the ionization parameter is 0.4~dex higher.

In Q1009$+$2956 we have the opposite situation, in which $W_{\rm
  Ly\alpha}$ is larger than $W_{\rm N\,V}$, i.e., \lya\ is
diluted more by emission line than \nv, suggesting that metallicity
may have been overestimated. To assess whether this is indeed the
case, we explored a new set of models in which the $C_{\rm e}$ has the
minimum value possible and $C_{\rm c}$ has the maximum value possible.
According to Eq.~(\ref{ccmax}) and (\ref{cemin}), these values are
$C_{\rm c}=1$ and $C_{\rm e}=0.245$. Thus, the effective covering
factors are $C_{\rm f}=0.593$ for \lya, and $C_{\rm f}=0.895$ for
\ciii. By setting $\log{\left(n_{\rm H}/{\rm cm}^{-3}\right)}=8$, we
find an acceptable model at $(\log{(Z/Z_\odot)},\log{U})=(2.5,0.4)$,
very similar to our previous best-fitting model which had $C_{\rm c} =
C_{\rm e}$.  

In the calculation above, we assumed the same {\it average} value of
$W$ of two members of the \nv\ doublets. However, the value of $W$
could differ between the blue and red members of the \nv\ doublet,
which could affect not only the metallicity estimate but also values
of the coverage fractions. In this particular case, however, the value
of $W$ is the same for the two members of the double, within
uncertainties.
\section{Discussion} \label{discussion}

\subsection{High Metallicities}

One of the main results of this work is that the metallicities in
intrinsic NAL systems are quite high, $\log(Z/Z_\odot)\gtrsim1$.  Several
possible scenarios have been proposed to produce super-solar
metallicities in the vicinity of the central massive black holes of
quasars \citep{ham99}. The most natural scenario involves normal
galactic chemical evolution \citep[see][]{ham92,ham93}. By considering
spectral synthesis and chemical enrichment models of the N{\,\sc
  v}/C{\,\sc iv} and N{\,\sc v}/He{\,\sc ii} emission line ratios, it
was suggested that nitrogen is over-abundant by factors of 2--9 in
high redshift ($z>2$) quasars \citep{ham99}.  Highly evolved gas with
$\log(Z/Z_\odot) > 0.5$--1.1 is needed to produce such a large
over-abundance of nitrogen.  This evolutionary model requires a large
number of high mass stars and implies a flat initial mass function
(IMF) and rapid star formation. The power-law index of the IMF
($\Phi\propto M^{\gamma}$) can be in the range $\gamma=0.9-1.4$ and
star formation must occur in less than $0.5$ Gyr for $z>4$ objects
\citep{ham92}. This vigorous star formation is consistent with that
found in the environment of active galactic nuclei (AGNs). The required time
scales, metallicities, and IMFs are similar to those of elliptical
galaxies and bulges of disk galaxies \citep{rom02}.  Quasars with more
massive host galaxies tend to be more metal rich, based on
observations of massive ellipticals \citep{cle08,del06}.  Therefore,
the high abundances derived from our absorbing systems are not
surprising, assuming this scenario. The quasars could undergo
extensive chemical evolution before they are observable.  Other
possible processes within the accretion disk that can lead to
supersolar metalicities include metal enrichment by star trapping
\citep{art92,shields96}, star formation \citep{col99,good04}, and
non-stellar nucleosynthesis \citep{kun96}. However, it is unlikely
that these mechanisms can produce metallicities as high as
$\log(Z/Z_\odot)\sim 1$.

Early measurements of extremely high quasar metallicities based on
BALs \citep{tur86} could have been overestimated because partial
coverage of the background emission was not taken into account,
leading to underestimation of column densities \citep{ham99}. However,
since our analysis of NALs takes partial coverage into account, our
measurements do not suffer from this bias.  Actually, super-solar
metallicities have also been derived for associated absorption systems
in low redshift AGNs \citep{pap00,fie05,gab06}. However, those
metallicities are not higher than $\log(Z/Z_\odot)\sim 1$, i.e.,
somewhat lower than our results for quasar intrinsic NAL systems. This
difference could be due to the higher redshifts of our NAL systems
and/or due to their presence in more luminous AGNs.  Rapid and
vigorous stellar evolution is likely to accompany quasar activity, as
suggested by a correlation between the Eddington ratio and the BELR
metal abundance spanning 3 orders of magnitude in both quantities
\citep[e.g.,][]{shem04}.  The resulting high-metallicity gas could
then be diluted with a low-metallicity interstellar medium as it moves
outward in the host galaxy. Eventually, it can escape the host-galaxy
potential, enriching the interstellar and intergalactic medium.

It is also possible that we are looking only at the tip of the
iceberg, i.e., at the most highly enriched gas. A diversity of absorber
properties, including metallicity, has been reported by
\citet{pet96}. Not only can absorbers have different coverage
fractions for the continuum and BELR but also different ions/species
can have different coverage fractions due to the multi-phase structure
of the gas.  Different coverage fractions for different ions/species
are also inferred from the general properties of the strong \civ\
family of intrinsic NALs \citep{mis07}, which usually have a saturated
\lya\ line, but no detected \nv, implying a lower ionization state for
the gas. In the strong-\nv\ NAL systems studied in this paper, the \ovi\
doublet may sometimes be under-produced by the models, which may be the
result of a greater coverage fraction of the \ovi\ absorber.  Thus the
high metallicities that we derived for the NAL gas in which \nv\ is
detected need not apply to all of the gas surrounding the central
engine.

\subsection{Properties of the Absorbing Gas}

The ionization parameters that we obtained are similar to those for
BALs. A typical value of a BAL ionization parameter is $\log{U}\sim0$
\citep[with a range of $-1.0$ to $+0.6$; e.g.,][]{cre03,wam95,tel98}.
Both an intrinsic NAL and a BAL are observed in the spectrum of
Q0059$-$2735 by \citet{wam95}, who show that both can be fitted with
$\log{U}=-0.7$. \citet{ham98} finds that $\log{U}\gtrsim -0.6$ is
required to model the BAL of PG1254$+$047, while \citet{tel98}, using
a two-slab geometry to model the BAL system of QSO~SBS~1542$+$541, find
$-1.1<\log{U}<0.6$ for the lower-ionization zone and
$\log{U}\gtrsim0.3$ for the higher-ionization zone.

If we assume that the outflow includes thin filaments (justified
below) that are responsible for the UV absorption lines that we
observe, we can estimate the outflow rate in these filaments as
\begin{equation}
\dot M = 4\pi r^2\left(\delta r\over r\right)\rho \vel
\left(\delta\Omega \over 4\pi\right) \; ,
\label{def_mdot}
\end{equation}
where $\rho$ is the typical mass density of absorbing filaments, $r$
is their typical distance from the center, $\delta r$ is their typical
thickness in the radial direction and $\delta\Omega$ is the solid
angle they subtend to the center of the flow.  We can re-cast the
above expression by using the definition of the ionization parameter,
$U\equiv Q/4\pi r^2 n_{\rm H} c$, to replace $4\pi r^2\rho$ ($Q$ is
the rate of emission of ionizing photons at $E>1$~Ry, and $n_{\rm H}$
is the hydrogen number density). We also note that the thickness of a
filament is related to the hydrogen density, and radial column
density, $N_{\rm H}$, via $N_{\rm H}=n\,\delta r$. Thus we can write
\begin{equation}
\dot M = 5.2\times 10^{-5}\; N_{18}\; \vel_2\; f_{-1} 
\left(Q_{58}\over n_6\; U\right)^{1/2}~{\rm M_\odot~yr}^{-1}\; ,
\label{calc_mdot}
\end{equation}
where $N_{\rm H} = 10^{18}\, N_{18}~{\rm cm}^{-2}$, $\vel=10^2\, \vel_2~{\rm
km~s}^{-1}$, $(\delta\Omega/4\pi)=10^{-1}\, f_{-1}$, $Q=10^{58}\,
Q_{58}~{\rm s}^{-1}$, and $n_{\rm H}=10^6\, n_6~{\rm cm}^{-3}$. We
also note that from the definition of the ionization parameter, we can
obtain an estimate of the distance of the absorbing filament from the
ionizing source.

To justify the assumption that the absorbers are thin, we note that
using the definition of $U$ and $N_{\rm H}=n\,\delta
r$, we obtain $\delta r/r = N_{\rm H}^2 4\pi c U/n_{\rm H} Q$.  The
values of $N_{\rm H}$ (neutral plus ionized) inferred from our
photoionization models are in the following range: $\log (N_{\rm
  H}/{\rm cm}^{-2}) \sim 16$--19 for HE0130--4021, $\log (N_{\rm
  H}/{\rm cm}^{-2}) \sim 17$--18 for Q1009+2956, and $\log (N_{\rm
  H}/{\rm cm}^{-2}) \sim 19$ for HS1700+6416, the values of $U$ are
given in Tables~\ref{tab-summary}--\ref{tab-bestq1700} and the values
of $Q$ are given in Table \ref{tab-sumsys}. Thus, any combination of
values of the above quantities yields $\delta r/r < 0.01$ which
indicates that the absorbers are geometrically thin along the line of
sight.

We can now use equation~\ref{calc_mdot} to obtain constraints on the
mass outflow rate. The necessary values of $N_{\rm H}$, $\vel$, $U$
and $Q$ are given in Tables~\ref{tab-summary}--\ref{tab-bestq1700}.
We also use a value of $f_{-1}\sim 5$ based on the results of
\citet{mis07}. Unfortunately, we have very poor constraints on the
hydrogen volume density, which lead to a very wide range of values for
the mass outflow rate, i.e.  $\dot M \sim {\rm few~M_\odot~yr}^{-1}$
for the lowest densities and $\dot M < 10^{-4}\; {\rm
  M_\odot~yr}^{-1}$ for $n_{\rm H} \geq 10^{10}~{\rm cm}^{-3}$. The
kinetic power of these filaments is negligible compared to the
electromagnetic luminosity ($\varepsilon_{\rm k}\equiv {1\over 2}\dot
M\, \vel^2/L_{\rm bol} \lesssim 10^{-5}$), however they may be
embedded in a hotter, more massive outflow or they may be accompanied
by a more massive outflow along a different line of sight.

Finally, we note that the poor constraints on the density also lead to
poor constraints on the distance of the filament from the ionizing
source, $r$. In Tables~\ref{tab-bestq1009}--\ref{tab-bestq1700} we
list the constraints on the radial distance of the filaments as a
function of density for Q1009+2956 and HS1700+6416 (the results for
HE0130--4021 are very similar). For $n_{\rm H} \geq 10^9~{\rm
  cm}^{-3}$, the absorbers turn out to be very close to the ionizing
source, at distances comparable to the size of the BELR, while at the
lowest densities, the filaments are located in the outskirts of the
host galaxy.
\section{Summary and Conclusions} \label{conclusion}

In this paper, we modelled the intrinsic NAL systems in the spectra of
the quasars HE0130$-$4021, Q1009$+$2956, and HS1700$+$6416, We
identify these systems as intrinsic because they exhibit partial
coverage in the \nvdblt\ doublet. Using the photoionization code
Cloudy, we simulate and reconstruct each absorption feature observed
in a system by adjusting the metallicity (relative to the Sun),
$Z/Z_\odot$, ionization parameter, $U$ and hydrogen volume density,
$n_{\rm H}$.  Our main conclusions are:

\begin{enumerate}

\item All three intrinsic systems have metallicities of
  $\log(Z/Z_\odot) \gtrsim 1$--2, regardless of constraints on $U$ or
  $n_{\rm H}$.  We find the high metallicities that we infer
  cannot alternatively be explained by different coverage fractions of
  the continuum source and BELR. The origin of these 
  supersolar metallicities is
  unknown, but we speculate that they may result from intensive star
  formation before the quasar becomes observable. They could also be
  related to an inhomogeneous metallicity distribution in the vicinity
  of the quasar central engine.

\item All three intrinsic systems require high ionization parameters,
  $\log{U}\sim0$, which are similar to those derived for BALs.

\item We can constrain the coverage fractions of the continuum source
  and the BELR separately, and find that the continuum source
  is not fully covered by the absorbers.

\item Although we cannot constrain the hydrogen volume density very
  well, we do find that the widths of the synthesized line profiles
  become larger than the observed ones at very high densities. This
  is a result of a sharp increase in temperature at high densities,
  produced by free-free absorption of infrared photons by free
  electrons. This leads us to conclude that the hydrogen volume
  density is in the range $2 \lesssim \log(n_{\rm H}/{\rm cm}^{-3})
  \lesssim 9$.

\item As a result of the large range of possible densities we cannot
  obtain robust estimates of the mass outflow rate in the filaments
  responsible for the \nv\ absorption. We can conclude, nevertheless,
  that $\dot M \lesssim {\rm few~M_\odot~yr}^{-1}$, and the
  corresponding power is a negligible fraction of the electromagnetic
  luminosity.

\item The strength of the \ovidblt\ doublet is underproduced by the
  models that reproduce the other absorption lines. This suggests that
  the \ovi\ lines are either contaminated by lines in the Lyman forest
  of that they arise in a separate gas phase with a larger coverage
  fraction.

\end{enumerate}

The strong-\nv\ absorbers studied in this paper make up one of two,
apparently distinct, families of intrinsic absorber identified by
\citep{mis07}. Our next goal is to study the physical properties of
the other family of absorbers, the strong-\civ\ family and
probe the origin of the two families.

{\acknowledgments We acknowledge helpful comments from an anonymous referee.
We thank David Turnshek for constructive
suggestions, and Richard Wade and Lijun Gou for useful discussions. We
also thank David Tytler and his group members for kindly providing us with
the spectra of the absorption systems used in this work. This work was
supported by NASA grant NAG5-10817 and by NSF grant AST-0807993.}

\end{document}